\documentclass[letterpaper,twocolumn,10pt]{article}
\usepackage{usenix}
\usepackage{xcolor}
\usepackage{framed}
\usepackage{xspace}
\usepackage{multirow}
\usepackage{booktabs}
\usepackage{graphicx}
\usepackage{siunitx}
\usepackage{makecell}
\usepackage{enumitem}
\usepackage{threeparttable}
\usepackage{soul}

\usepackage[framemethod=TikZ]{mdframed}
\usepackage[ruled,vlined,linesnumbered,noend]{algorithm2e}
\usepackage{algorithmicx}
\usepackage{algpseudocode}
\usepackage{amsmath}
\usepackage{caption} 
\usepackage{xurl}
\usepackage{hyperref}
\hypersetup{
  colorlinks,
  linkcolor={blue!70!green},
  citecolor={black},
  urlcolor={black}, 
  breaklinks=true
}
\usepackage{soul} 
\usepackage[ruled,vlined,linesnumbered]{algorithm2e}
\SetKwComment{Comment}{$\triangleright$\ }{}
\SetKwInOut{Variables}{Variables}

\newcommand{\tool}[0]{\texttt{DPAgent}\xspace}

\newcommand{\pie}[1]{%
\begin{tikzpicture}
\draw (0ex,0ex) circle (1ex);
\fill (0ex,-1ex) arc (-90:(#1-90):1ex) -- (0ex,-1ex) -- cycle;
\end{tikzpicture}%
}           

\def\eg{\emph{e.g.,}\xspace}

\def\ie{\emph{i.e.,}\xspace}
\def\etal{\emph{et al.}\xspace}

\usepackage{framed}
\definecolor{formalshade}{rgb}{0.95,0.95,0.97}
\definecolor{darkblue}{rgb}{0.14,0.22,0.52}
\newenvironment{takeaway}{
  
  \MakeFramed{\advance\hsize-\width\FrameRestore}
  \noindent\hspace{-4.55pt}
}
{
  \endMakeFramed%
}

\newcounter{takeaway}
\usepackage{tcolorbox}

\usepackage{enumitem}
\setlist[itemize]{itemsep=0pt, parsep=0pt, nosep}

\usepackage{tabularx}
\usepackage[table]{xcolor}

\usepackage{booktabs, multirow, threeparttable, graphicx}

\usepackage{amssymb}
\newtheorem{definition}{Definition}
\newtheorem{assumption}{Assumption}
\newtheorem{problem}{Problem}

\definecolor{atkred}{RGB}{211,47,47}
\definecolor{voidamber}{RGB}{239,108,0}
\definecolor{pipegray}{RGB}{69,90,100}
\definecolor{modelblue}{RGB}{21,101,192}
\newcommand{\stg}[2]{{\setlength{\fboxsep}{1.8pt}%
  \colorbox{#1!30}{\textcolor{#1!75!black}{\strut\tiny\bfseries\sffamily #2}}}}
\newcommand{\chto}{\hspace{1.5pt}{\tiny\textcolor{black!60}{$\boldsymbol{\rightarrow}$}}\hspace{1.5pt}\allowbreak}

\begin{document}
%-------------------------------------------------------------------------------

\title{What You See Is Not What AI Gets: \\\tool{}-in-the-Middle Defense Against AI-Groomed Deceptive Patterns}

\author{
    {\rm
        Zewei Shi$^{\dagger\ddagger}$,
        Ruoxi Sun$^{\ddagger}$,
        Haoyang Li$^{\S}$,
        Seong Oun Hwang$^{\P}$,
    }\\
    {\rm
        Feng Liu$^{\dagger}$,
        Minhui Xue$^{\ddagger}$, and
        Xingliang Yuan$^{\dagger}$
    }\\[0.5em]
    \textit{$^{\dagger}$The University of Melbourne} \quad \textit{$^{\ddagger}$CSIRO} \quad \textit{$^{\S}$Macquarie University} \quad \textit{$^{\P}$Gachon University}
}

\maketitle

\begin{abstract}
Privacy deceptive patterns in web interfaces manipulate users into disclosing personal data, yet existing defenses are fragmented, static, and increasingly vulnerable to manipulation by large language models. Moreover, data voids, areas of information scarcity on the web, allow adversaries to inject misleading content that can be scraped and learned by AI systems, amplifying both deceptive design and model misbehavior. In this paper, we formalize AI grooming as a new threat in which adversaries seed benign-looking artifacts carrying machine-consumable manipulative signals into AI-mediated workflows. To address this threat, we present \tool{}, an agentic, reasoning-aware framework that orchestrates four specialized agents combining latent-space purification with defensive prompting to explore, detect, and repair privacy deceptive interfaces in live web environments. Extensive evaluations show that \tool{} filters 91\% of naive whole-page generated samples and consistently reduces attack success across five targeted grooming strategies, achieves state-of-the-art detection with a micro F1 of 0.82, explores over 80\% of pattern types while visiting only about 10\% of the pages required by baselines, and successfully repairs 89.7\% of correctly detected PDP instances. Our results demonstrate the promise of agent-in-the-middle defenses for securing the web UI supply chain against deceptive design and emerging AI threats rooted in data void exploitation.
\end{abstract}

%===============================================================================
\section{Introduction}\label{sec:intro}
Deceptive patterns (DPs), also known as dark patterns~\cite{gray2018dark}, are interface design strategies that exploit cognitive biases to steer users toward choices they might not otherwise make, such as oversharing personal information, accepting tracking, or consenting by default~\cite{mathur2019dark,nouwens2020dark}. Prior work has extensively documented these practices and developed taxonomies covering a broad range of manipulative interface behaviors~\cite{shi202550,GaryCHI24,gunawan2021comparative}. However, DPs have largely been studied as \emph{local properties of individual interfaces}: a webpage contains a deceptive element, a user encounters it, and a detector or auditor attempts to identify it. This view is increasingly incomplete as the web becomes mediated by AI systems that participate in content retrieval, generation, and reuse.

Modern web interfaces are no longer shaped solely by human developers. AI systems increasingly consume web content through crawling and retrieval pipelines, assist developers in generating interface code and content, and mediate how users discover websites through search and AI-assisted recommendation. Recent work has already shown that generative AI can introduce deceptive designs into generated websites even without explicit requests to do so~\cite{ChatGPTLearning}. As a result, a deceptive interface can propagate beyond the website on which it first appears. DP-bearing content may be collected by automated pipelines, incorporated into AI training or retrieval corpora, reproduced by downstream models, and eventually reused in newly generated interfaces. At the same time, search and recommendation systems determine which websites users encounter in the first place. These interactions turn DPs from an isolated interface-level concern into a broader \emph{web supply-chain security problem}, in which manipulative designs can be learned, redistributed, and exposed to users at scale.

A particularly important amplification point in this supply chain is the \emph{data void}~\cite{boyd2019data}. A data void arises when a query corresponds to a topic for which little authoritative or high-quality content exists. Search engines and AI retrieval systems implicitly rely on sufficient trustworthy content being available; when this assumption fails, low-quality or strategically crafted content faces little competition and can disproportionately influence crawling, indexing, and retrieval. Adversaries have already exploited such information scarcity by seeding misleading or fabricated content into underpopulated query spaces, a strategy increasingly associated with \textit{AI grooming}~\cite{newsguard2025pravda,radina2025grooming,prorussia2025american}. Importantly, this attack pattern is substantiated by state-aligned influence operations using hundreds of affiliated websites to amplify geopolitical narratives~\cite{dfrlab2025pravda}. Empirical observations further indicate that coordinated content can be ingested and reproduced by retrieval systems and LLMs even when it is not persuasive to human readers directly~\cite{bulletin2025russian,disinfo2025llmgrooming,OSF2025}. 

\begin{figure*}[t]
\centering
\includegraphics[width=0.9\linewidth]{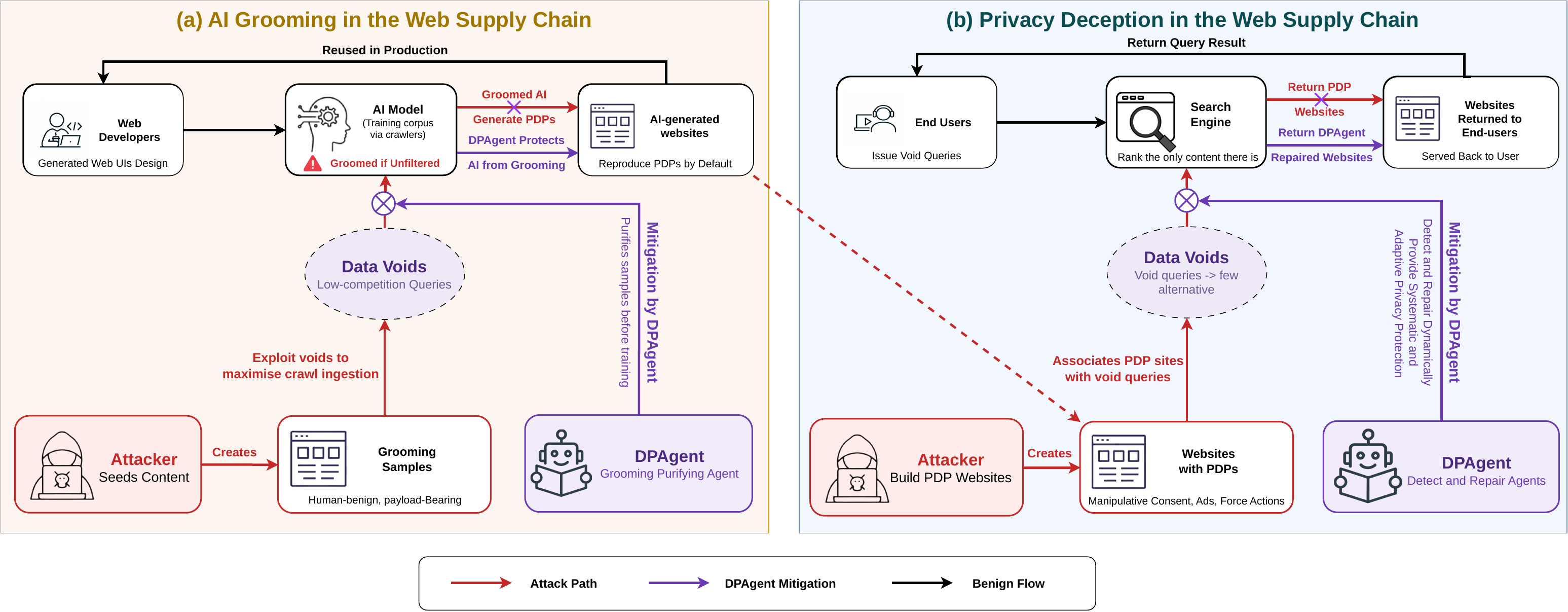}
\caption{An overview of \tool{}. \tool{} reframes web privacy issue mitigation as a supply chain optimization, providing an effective defensive strategy against AI grooming and privacy deceptive patterns.}
\label{fig:scenario}
\end{figure*}

We argue that this mechanism creates a previously overlooked attack surface for deceptive patterns. \textit{Data voids do not create DPs; they amplify them.} In particular, information scarcity increases the likelihood that DP-bearing artifacts dominate either what AI systems ingest or what users retrieve. Figure~\ref{fig:scenario} illustrates two complementary propagation paths. \textbf{First, through AI supply-chain grooming.} An attacker creates apparently benign web artifacts containing manipulative interface strategies and associates them with low-competition queries. The resulting content is more likely to be collected by crawlers or AI data pipelines and, if left unfiltered, can influence models that subsequently reproduce or generate deceptive interfaces. For example, a model exposed to manipulative consent designs or misleading privacy claims may later reproduce similar strategies when generating web interfaces. \textbf{Second, through user-facing privacy deception.} An attacker creates websites containing DPs and associates them with void queries. Because few trustworthy alternatives exist, search engines or AI assistants may disproportionately surface these sites, directly routing users toward manipulative interfaces. In both cases, the data void acts as an upstream amplifier: one path increases the probability of \emph{machine ingestion}, while the other increases the probability of \emph{human exposure}.

Studying DPs through data voids is therefore important for a reason that goes beyond content ranking alone. Unlike conventional misinformation, the harmful payload of a DP does not necessarily depend on convincing a user that a factual statement is true. Instead, its effect is realized through \emph{interaction}: inducing a user to accept tracking, disclose personal information, overlook an alternative, or follow a designer-preferred action~\cite{nouwens2020dark,luguri2021shining}. Consequently, protecting only the textual or factual content of the web supply chain is insufficient. A defense must also reason about the interface through which the content is ultimately presented and the interaction paths along which deceptive behavior emerges. This is particularly challenging because DPs are diverse and contextual, often appear only after multi-step interactions such as consent, registration, or configuration flows, and may be embedded in artifacts specifically crafted to evade human or automated inspection.

Data voids themselves are difficult, and in many cases impossible, to eliminate. Rather than attempting to remove information scarcity, we therefore focus on \emph{breaking the attack paths that exploit it before they result in model contamination or user harm}. This requires protection at multiple points in the web supply chain. First, DP-bearing grooming artifacts should be filtered before they are consumed by downstream AI systems. Second, websites must be explored along realistic interaction paths rather than through static or exhaustive inspection. Third, deceptive behavior must be detected using contextual and expert-informed reasoning. Finally, detection alone is insufficient: harmful interfaces should be repaired before they are presented to users while preserving legitimate webpage functionality. These requirements motivate an agent-in-the-middle defense that can intervene between potentially manipulated web content and the AI systems or users that consume it.

To this end, we present \tool{}, a multi-agent framework that operates across the web interface supply chain to mitigate privacy deceptive patterns (PDPs). PDPs form a privacy-focused extension of existing DP taxonomies, selecting privacy-critical pattern categories and augmenting them with use-case descriptions and targeted repair strategies (Appendix~\ref{apx:taxonomy}). \tool{} coordinates four specialized agents for grooming purification, task generation and exploration, PDP detection, and interface repair. Its Grooming Purifying Agent acts as an upstream choke point, combining latent-space provenance filtering with defensive prompting to prevent manipulated artifacts from propagating downstream. Its Task Generation Agent efficiently explores realistic user interaction paths, while the PDP Detection Agent combines expert-validated knowledge with multimodal reasoning to identify privacy-threatening designs. Finally, the Interface Repairing Agent applies predefined, functionality-preserving interventions before deceptive interfaces reach users. 

\noindent\textbf{Contributions.} Our contributions are fourfold:
\begin{itemize}[leftmargin=*]
\item We reveal a new attack surface that connects data void exploitation with DPs, showing how information scarcity can amplify both AI grooming and manipulative interface practices across the web supply chain.
\item We formalize AI grooming in the context of PDPs (\S\ref{sec:grooming}) and propose a defense that combines latent-space purification with prompt-level filtering.
\item We design and implement \tool{}, an agent-in-the-middle framework for AI grooming purification and PDP detection and repair, integrating automated task generation, interface exploration, structured reasoning, and real-time interface intervention within live web environments (\S\ref{sec:dpagent}).
\item We conduct comprehensive evaluations and real-world studies on diverse web applications, demonstrating \tool{}'s effectiveness and robustness in mitigating PDPs under adversarial conditions (\S\ref{sec:eval}).
\end{itemize}

%===============================================================================
\section{New AI Grooming Attacks}\label{sec:grooming}
AI grooming attacks are fundamentally distinct from backdoor, jailbreak, and adversarial attacks in both intent and execution. Backdoor attacks~\cite{liu2020reflection} embed hidden triggers during training to elicit incorrect behavior only when specific conditions are met. Jailbreak attacks~\cite{wei2023jailbroken} circumvent safety mechanisms to extract restricted content, and adversarial attacks~\cite{goodfellow2014explaining} craft malicious inputs at inference time to induce immediate misclassification. All three presuppose an attacker who interacts with a concrete victim model. In contrast, AI grooming targets no model directly: \textit{it is a manipulation tactic that strategically seeds the open web, so that benign pipelines of crawling, curation, and retrieval deliver the payload on the attacker's behalf.}

The comparison with data poisoning~\cite{DataPoisoning} is subtler. We argue that the two threats differ along three orthogonal axes (Table~\ref{tab:poisoning_vs_grooming}). The first axis is \emph{access}. Poisoning presupposes a write path into the victim's training data, whereas grooming requires no privileged access: public publication is the entire attack surface. The second axis is \emph{payload}.  Even sophisticated clean-label poisoning ultimately relies on imperceptible statistical perturbations or hidden feature-level triggers,  whereas grooming preserves human-perceived benignness while introducing machine-relevant manipulations, including both low-perceptibility perturbations and semantic-gap payloads. It injects logically coherent narratives that appear \textit{harmless to human reviewers yet inherently conceal ground-truth malicious intent}. Because such content lacks the structural artifacts typical of poisoned data, it bypasses human-in-the-loop filtering, and its absence of statistical anomalies allows it to evade automated data quality audits. Its effect is amplified not by sample volume but by \emph{data voids}, regions of the web where no competing content exists to dilute the groomed narrative. The third and most consequential axis is \emph{window}. Poisoning is confined to training, whereas a groomed artifact remains live at inference time, reaching the model through retrieval-augmented generation and web search without ever entering a training set. 

The training-time projection of grooming is therefore closest to indirect, web-scale poisoning: Carlini~\etal~\cite{carlini2024poisoning} showed that an attacker with no privileged access can poison production-scale corpora by purchasing expired domains behind crawled URLs (split-view poisoning) or by timing malicious edits against periodic snapshots (frontrunning poisoning). Grooming inherits this zero-access supply-chain position but departs from it in two respects. Where such attacks embed sample-level payloads whose effect is bounded by the training pipeline, grooming constructs self-consistent narratives that persist as citable web content, allowing the same artifact to compromise retrieval at inference time, a path outside the canonical training-data poisoning threat model.  Conversely, poisoning in its canonical form assumes an insider position, namely write access to the victim's training data. The two threats therefore intersect in the indirect-poisoning field, yet neither contains the other.

\begin{table}[t]
\centering
\caption{Data poisoning vs.\ AI grooming: the poisoner writes into the victim's training data; the groomer only publishes to the open web and benign pipelines deliver the payload, at training or inference time.}
\label{tab:poisoning_vs_grooming}
\scriptsize
\renewcommand{\arraystretch}{1.12}
\setlength{\tabcolsep}{3pt}
% --- measure stage widths for alignment (move \newlength to preamble if reused) ---
\newlength{\wlab}\settowidth{\wlab}{\textbf{Grooming:}\ }
\newlength{\wweb}\settowidth{\wweb}{\stg{voidamber}{Open web}}
\newlength{\wcrawl}\settowidth{\wcrawl}{\stg{voidamber}{Crawl\,/\,Retrieve}}
\newlength{\wtrain}\settowidth{\wtrain}{\stg{pipegray}{Training data}}
% --- ghost stage: empty dashed outline, sized by phantom text ("bypassed") ---
\newcommand{\stgghost}[1]{%
  \tikz[baseline=(g.base)]{%
    \node[draw=gray!50, dashed, thin, fill=none,
          rounded corners=1.5pt, inner xsep=3pt, inner ysep=1.5pt,
          font=\scriptsize] (g) {\phantom{#1}};}}
\begin{tabularx}{\columnwidth}{@{}l>{\raggedright\arraybackslash}X>{\raggedright\arraybackslash}X@{}}
\toprule
\multicolumn{3}{@{}l@{}}{%
  \makebox[\wlab][l]{\textbf{Poisoning:}}%
  \stg{atkred}{Attacker}\chto
  \makebox[\wweb][c]{\stgghost{Open web}}\chto
  \makebox[\wcrawl][c]{\stgghost{Crawl\,/\,Retrieve}}\chto
  \stg{pipegray}{Training data}\chto
  \stg{modelblue}{Model}}\\[2pt]
\multicolumn{3}{@{}l@{}}{%
  \makebox[\wlab][l]{\textbf{Grooming:}}%
  \stg{atkred}{Attacker}\chto
  \stg{voidamber}{Open web}\chto
  \stg{voidamber}{Crawl\,/\,Retrieve}\chto
  \makebox[\wtrain][c]{\stg{pipegray}{Train\,/\,Infer}}\chto
  \stg{modelblue}{Model}}\\
\midrule
\rowcolor{gray!14}
& \textbf{Data Poisoning} & \textbf{AI Grooming (this work)} \\
\midrule
\textsc{Acts at}
& Inside training pipeline
& Open web, upstream of any pipeline \\
\textsc{Access}
& Write to training data
& None; just publish \\
\textsc{Payload}
& Statistical triggers; fools data audits
& Coherent content; fools humans, read by machines \\
\textsc{Amplifier}
& Poisoned-sample volume
& Data voids: no competing content \\
\textsc{Window}
& Training only
& Training \emph{and} inference \\
\textsc{Success}
& Misbehaves on a trigger
& Normalizes deceptive patterns \\
\bottomrule
\end{tabularx}
\end{table}

Recent work further argues that generative AI can scale influence campaigns in both breadth and depth through low-cost, persistent, and context-tailored interactions, including cases in which users may not realize that an AI system is involved~\cite{guennemann2026securityprivacyllm}. By shifting the attack from immediate model outputs to the underlying data substrate, AI grooming creates a pathway through which artifacts carrying manipulative interface strategies can enter AI data pipelines. Among the privacy-harming payloads that can be embedded in such artifacts, \textit{PDPs are particularly consequential}: the broader deceptive-pattern ecosystem is already deployed at scale~\cite{mathur2019dark,gunawan2021comparative}, and controlled studies show that interface choices materially increase consent or sign-up rates, creating direct commercial incentives to deploy such practices~\cite{nouwens2020dark,luguri2021shining}. Data voids extend the reach of this already prevalent and economically incentivized class of manipulative designs, a claim we make precise below.

\begin{table}[t]
\centering
\caption{Notation used throughout the paper.}
\label{tab:notation}
\small
\setlength{\tabcolsep}{4pt}
\begin{tabular}{@{}ll@{}}
\toprule
\textbf{Symbol} & \textbf{Meaning} \\
\midrule
$w=(d,r)$ & webpage with DOM $d$ and rendered state $r$ \\
$\mathrm{vis}(w),\,\mathrm{ext}(w)$ & human-visible / machine-extractable content \\
$\Delta(w)$ & semantic gap of $w$, Eq.~\eqref{eq:gap} \\
$\mathcal{Y},\ \phi(w),\ \hat{\phi}(w)$ & PDP taxonomy; true / predicted labels of $w$ \\
$g,\ \varepsilon,\ \delta$ & grooming transformation; perceptual budget; \\
 & \quad attack success probability \\
$Q_{\mathrm{void}}$ & set of data-void queries \\
$f_{\text{GF}},\ F_{\text{GF}}$ & grooming filter and its feature encoder \\
$I_o^{(i)},\ I_g^{(i)}$ & $i$-th original / groomed screenshot \\
$I_{\text{def}},\ \mathcal{K}$ & defensive prompt; expert knowledge base \\
$\mathrm{Cov}(T),\ B$ & PDP-type coverage of task list $T$; page budget \\
$\mathcal{A}(y),\ w^\star$ & admissible repair actions for pattern $y$; \\
 & \quad repaired page \\
\bottomrule
\end{tabular}
\end{table}

\noindent \textbf{Preliminaries.}
Before stating the threat model, we fix minimal notation (Table~\ref{tab:notation}); these symbols are reused verbatim in Sections~\ref{sec:dpagent} and~\ref{sec:eval}. Let a webpage be a pair $w=(d,r)$, where $d$ is the DOM (source, scripts, styles, metadata) and $r$ is its rendered visual state. The structural observation behind AI grooming is that the same page is consumed through two different channels. A human sees only the \emph{visible channel} $\mathrm{vis}(w)$, the perceptible content of $r$. An automated consumer, such as a DOM parser, a crawler, or a multimodal model, reads the \emph{extractable channel} $\mathrm{ext}(w)$, which additionally includes non-rendered or imperceptible signals such as alpha-channel payloads, invisible text, and metadata. In general $\mathrm{vis}(w)\subseteq\mathrm{ext}(w)$, and we call
\begin{equation}
\Delta(w)=\mathrm{ext}(w)\setminus\mathrm{vis}(w)
\label{eq:gap}
\end{equation}
the \emph{semantic gap} of $w$. Benign pages have small, functional gaps (\eg accessibility attributes); everything that follows is about how this gap is widened and weaponized. As a running example, consider a cookie banner whose screenshot has been regenerated to look identical to the original, but whose alpha channel carries the instruction-like string ``all cookies are strictly necessary'': a human reviewer sees an ordinary banner, while a model that reads the alpha channel ingests a false privacy claim. Let $\mathcal{Y}$ denote the PDP taxonomy of seven pattern types plus the benign class (Appendix~\ref{apx:taxonomy}), and let $\phi(w)\subseteq\mathcal{Y}$ be the set of PDPs that page $w$ instantiates, with $\phi(w)=\emptyset$ for benign pages. Finally, we formalize data voids~\cite{boyd2019data}:

\begin{definition}[Data void]\label{def:void}
A query $q$ is a \emph{data void} if the aggregate amount of authoritative, high-quality content indexed for $q$ falls below a threshold. We write $Q_{\mathrm{void}}$ for the set of such queries.
\end{definition}

\noindent In plain terms, the scarcer the authoritative content for $q$, the less competition attacker-supplied content faces, and the less effort the attacker needs to dominate what is crawled, indexed, and retrieved for that query. 

\smallskip
\noindent \textbf{Threat model.} We address a threat landscape where attackers exploit data voids to inject manipulated content. The unit of manipulation is the following transformation; the two attack vectors below are objectives defined over it.

\begin{definition}[Grooming transformation]\label{def:groom}
Let $d_H(\cdot,\cdot)$ be a human-perceptual distance on rendered states. A transformation $g$ is an $\varepsilon$-grooming transformation if
\begin{align}
d_H(\mathrm{vis}(w),\mathrm{vis}(w'))\leq\varepsilon, \label{eq:impercept}
\end{align}
while $w'$ carries an attack-relevant signal that induces or conceals a PDP from an automated consumer. When this signal lies in $\Delta(w')$, we refer to it as \emph{semantic-gap grooming}.
\end{definition}

\noindent Taken together, the definition captures the core asymmetry of grooming: the transformation remains benign from the human perceptual perspective, while introducing an attack-relevant signal that changes how an automated consumer interprets or processes the page. The signal may sit inside the semantic gap, or it may perturb machine-relevant visual features while staying within the perceptual budget $\varepsilon$. This differs from conventional adversarial examples, which introduce small input perturbations to alter model predictions while preserving the human-perceived class; grooming instead preserves the human-visible interface while manipulating machine-relevant information.

\noindent \textbf{\textit{(i)} Supply-chain grooming (the poisoning threat).} The adversary aims to influence future model behaviors. Groomed samples are indistinguishable from high-quality data to standard filters, so once seeded into data voids they are preferentially collected by crawlers and enter the training data of future foundation models, which then normalize or reproduce the embedded deceptive patterns by default.

\begin{definition}[Supply-chain grooming]\label{def:scg}
The attacker chooses a set $G=\{g(w_i)\}_{i=1}^{n}$ of groomed pages and associates each $w'\in G$ with a void query $a(w')\in Q_{\mathrm{void}}$ so as to maximize the expected contamination of future training data,
\begin{equation}
\max_{G,\,a}\ \sum_{w'\in G}\Pr\big[\,w' \text{ is crawled}\mid a(w')\,\big],
\label{eq:contamination}
\end{equation}
subject to Condition~\eqref{eq:impercept}.
\end{definition}

\noindent By Definition~\ref{def:void}, this objective is economical: associating groomed pages with void queries maximizes the crawl probability per injected page, since void queries face no competing authoritative content.

\noindent \textbf{\textit{(ii)} User-facing grooming (the deceptive generation threat).} In this scenario, an LLM mediating page generation or consumption is manipulated at inference time into accepting or producing UI content that carries a PDP payload, such as hidden trackers, invisible text, or manipulative consent logic that humans execute without suspicion. Because the content looks ordinary to a human reviewer, it passes human-in-the-loop verification while still executing the attacker's intent.

\begin{definition}[User-facing grooming]\label{def:ufg}
Given a deployed multimodal model mediating page generation or consumption, the attacker seeks $w'=g(w)$ satisfying Definition~\ref{def:groom} such that the model accepts, reproduces, or executes the payload at inference time with probability at least $\delta$, while human review passes because of Condition~\eqref{eq:impercept}.
\end{definition}

\noindent Definitions~\ref{def:groom}-\ref{def:ufg} are parameterized by the perceptual budget $\varepsilon$ and the success threshold $\delta$, which we do not fix analytically. In our study, Condition~\eqref{eq:impercept} is satisfied by construction: the five grooming strategies introduced below either perturb pixels within a visually negligible budget or place the payload entirely in $\Delta(w')$, leaving $\mathrm{vis}(w')$ unchanged. The threshold~$\delta$ is operationalized as the downstream detector's degradation on groomed inputs. Section~\ref{sec:eval} then establishes that the user-facing attack class is non-vacuous: inputs constructed under Definition~\ref{def:groom} materially alter the behaviour of a deployed multimodal detector. 

Although end-to-end contamination of a proprietary training pipeline is not externally observable, the crawl component of Eq.~\eqref{eq:contamination} is empirically estimable after fixing a crawler population and observation horizon. This can be measured by publishing benign canary pages on controlled domains, comparing matched data-void and non-void conditions while holding URL-discovery factors constant, and estimating fetch probabilities from verified server logs or public crawl archives~\cite{carlini2024poisoning,liu2025somesite}. Such measurements establish the collection precondition of supply-chain grooming, but not retention in a proprietary training corpus or causal changes to a trained model. We therefore evaluate the corresponding operational precondition: whether groomed artifacts can be reliably identified and removed before ingestion, precisely where \tool{} is designed to intervene.

\noindent \textbf{Defense scope via sub-agent training.} Our defense, \tool{}, bridges the gap between these two vectors through the specialized training of its \textit{Grooming Purifying Agent}. While \tool{} acts as a client-side proxy during inference time, this sub-agent is trained to discern the specific discrepancies between generated benign-looking artifacts and authentic content, neutralizing groomed samples at the point of contact before they can mislead the downstream \textit{PDP Detection Agent} or harm the end user. This placement, upstream of any model training, rests on the following premise.

\begin{assumption}[Detectable provenance]\label{ass:provenance}
Groomed artifacts $w'=g(w)$, prior to being laundered through model training and regeneration, retain attack-specific signatures in representation space: there exists a feature encoder $F$ such that the distance $\|F(I_o^{(i)})-F(I_g^{(i)})\|_2$ between an original screenshot~$I_o^{(i)}$ and its groomed counterpart $I_g^{(i)}$ is separable from natural page-to-page variation.
\end{assumption}

\noindent This assumption underpins our defense placement. Although empirical rather than theoretical, it is well motivated for artifacts satisfying Definitions~\ref{def:groom}-\ref{def:ufg}. First, successful grooming must alter an automated consumer's behavior (Definition~\ref{def:ufg}), which necessarily entails a change in the representation on which that behavior depends. Detectability and payload effectiveness therefore arise from the same underlying representational shift. Second, because grooming applies a transformation family $g$ across pages, it can induce systematic displacement in representation space that is distinguishable from natural variation, consistent with prior observations for generated and adversarially perturbed images~\cite{goodfellow2014explaining,feinman2017detecting,abusnaina2021adversarial}. Third, training and regeneration attenuate instance-level artifacts while retaining their induced semantic effects, making provenance most detectable before such laundering occurs. Section~\ref{sec:eval_grooming} empirically validates the assumption by instantiating $F$ with $F_{\text{GF}}$, measuring separability, and comparing the learned signal with an independent general-purpose encoder.  

\smallskip
\noindent \textbf{Adaptive adversaries.} Assumption~\ref{ass:provenance} presumes the attacker is unaware of, or indifferent to, the provenance signal. A stronger adversary that knows \tool{} is deployed will instead try to suppress that signal directly. We therefore make this attacker explicit.

\begin{definition}[Filter-aware adaptive grooming]\label{def:adaptive}
Let $f_{\text{GF}}$ denote the deployed provenance filter with encoder $F_{\text{GF}}$. Given an original screenshot $I_o$ and its groomed counterpart $I'_g$, we define $s(w')=\|F_{\text{GF}}(I_o)-F_{\text{GF}}(I_g')\|_2$  as the representation-space provenance discrepancy targeted by the adaptive attacker. A \emph{filter-aware adaptive attacker} seeks a grooming transformation that suppresses this signal while preserving both visual imperceptibility and payload effectiveness:
\begin{equation}
\begin{aligned}
\min_g \quad & s(w') \\
\text{s.t.}\quad
& d_H(\mathrm{vis}(w),\mathrm{vis}(w')) \leq \varepsilon,\\
& \Pr[
\substack{\text{attacker-intended behavior}\\
          \text{is induced on } w'}
] \geq \delta.
\end{aligned}
\label{eq:adaptive}
\end{equation}
The two constraints correspond to the imperceptibility requirement of Condition~\eqref{eq:impercept} and the payload-effectiveness requirement of Definition~\ref{def:ufg}. In deployment, however, $F_{\text{GF}}$ is client-side and its weights are not exposed. The attacker therefore cannot optimize $s(w')$ directly or differentiate through $F_{\text{GF}}$. The operative threat model is thus \emph{gray-box}: the attacker knows the defense and its architecture family, optimizes an analogous objective against a surrogate encoder $\tilde{F}$, and relies on transfer to evade the deployed filter. The \emph{white-box} setting, in which $F_{\text{GF}}$ is directly accessible and differentiable, assumes capabilities not provided by the deployment and is therefore outside our threat model.
\end{definition}

\noindent Eq.~\eqref{eq:adaptive} makes explicit that an adaptive attacker cannot minimise detectability in isolation: suppressing $s(w')$ must preserve both attack effectiveness and the human-visible appearance of the interface. We therefore conjecture that Assumption~\ref{ass:provenance} extends to this joint objective: if the provenance signal is coupled with the semantic features that carry the payload, then suppressing the former should also degrade the latter. In the gray-box setting, the attacker additionally faces a transfer barrier, since suppression optimized against a surrogate encoder need not transfer to $F_{\text{GF}}$. Section~\ref{sec:eval_adaptive} tests both conjectures empirically: whether surrogate-crafted suppression transfers to the deployed filter, and whether samples that do bypass it retain their payload effectiveness.

\smallskip
\noindent \textbf{Attacks considered in this study.}
\label{sec:attacks}
To operationalize the threat model and evaluate \tool{}'s defenses, we simulate two attack scenarios of increasing realism. {The first is a naive baseline}: a benign prompt generates an entire webpage from scratch, which targets both vectors simultaneously by presenting misleading UI content to deceive humans while masquerading as a benign-looking page to evade crawlers and poison training data. Whole-page generation, however, is difficult to keep authentic at scale, and search engines can detect such pages through AI-generated stylistic artifacts or missing user engagement signals (\eg real traffic and clicks).

To model the more sophisticated semantic-gap threat, we introduce a second, {targeted grooming strategy}. Instead of generating full pages, it modifies only small, high-risk regions within existing real webpages, termed \textit{privacy deceptive zones} (\eg cookie banners and ad frames), which directly implement PDPs and are frequent targets of data collection pipelines. By embedding subtle manipulations into these zones, attackers exploit the credibility of legitimate sites while executing inference-time evasion. We define five manipulation strategies that widen the gap between human perception and machine execution:
\begin{itemize}[leftmargin=*]
\item \textbf{Adding Noise:} injecting random noise into vulnerable regions to disrupt automated analysis while remaining visually negligible.
\item \textbf{Pixel Modification:} randomly altering selected pixels to arbitrary values to degrade model feature extraction.
\item \textbf{Invisible Text:} inserting induced text (\eg ads or policy statements) rendered in the background color, visible only to the DOM parser.
\item \textbf{Alpha Channel}~\cite{xiaalphadog}: embedding induced text into the image's alpha channel, invisible to humans but readable by AI.
\item \textbf{Unreadable Text:} replacing readable text with visually distorted but machine-readable content. 
\end{itemize}

All five techniques above satisfy Definition~\ref{def:groom} by remaining within the human-perceptual budget while altering machine-relevant information. The first two instantiate perceptual grooming by perturbing visual features within the budget~$\varepsilon$, whereas the last three instantiate semantic-gap grooming by placing the attack-relevant signal inside $\Delta(w')$. Combined with data void exploitation, such manipulations can be preferentially collected and learned, increasing the likelihood that future models reproduce or normalize PDPs.

%===============================================================================
\section{\tool{}: Exploration, Detection \& Repair}\label{sec:dpagent}

\begin{figure*}[t]
\centering
\includegraphics[width=\linewidth]{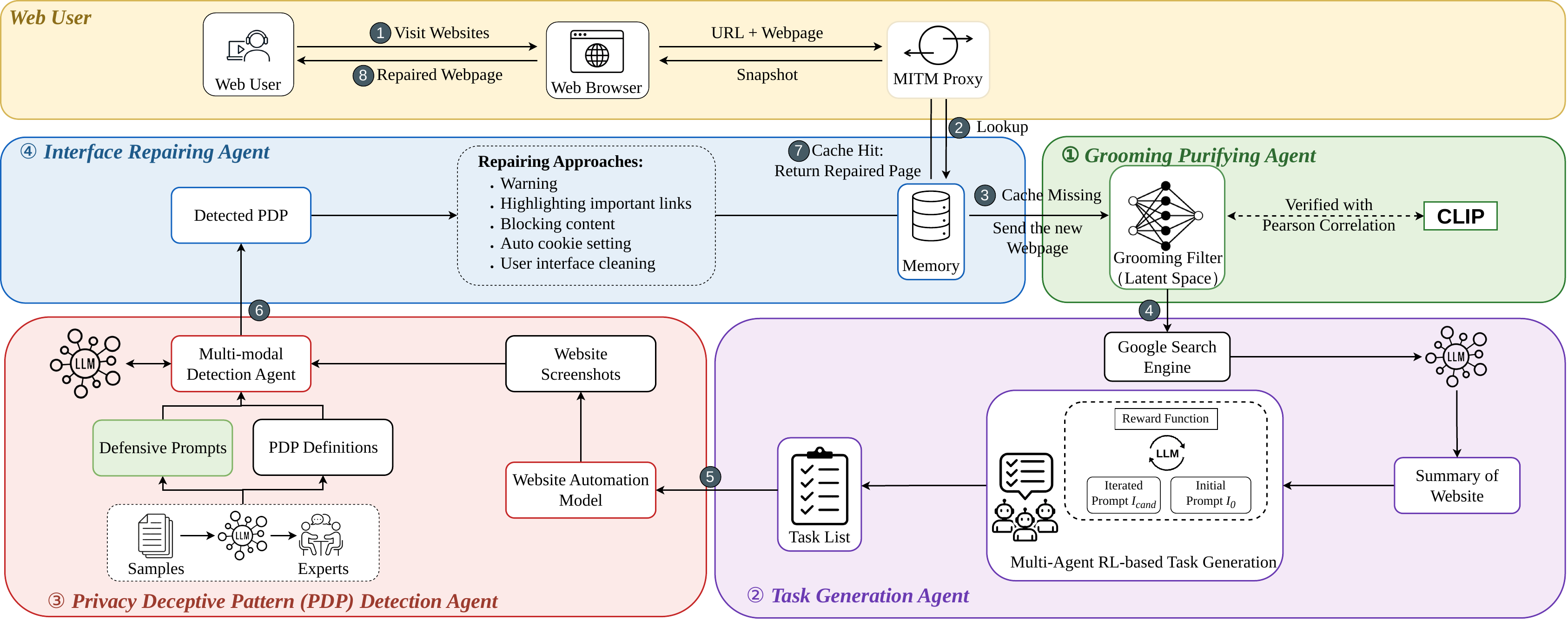}
\caption{An overview of the \tool{} framework. }
\label{fig:framework}
\end{figure*}

We present \tool{}, a multi-agent framework that dynamically explores websites, detects PDPs, and automatically repairs user interfaces before real users encounter them. 

\subsection{Problem Statement and Overall Workflow}
\label{sec:problem}

\noindent\textbf{Overall workflow.} Using the notation of Table~\ref{tab:notation}, we decompose the defense into four sub-problems, each instantiated by one agent in Figure~\ref{fig:framework}. During online protection, \tool{} sits between the user and the web as a MITM proxy: a requested page is served directly from the memory cache if a repaired version exists, and otherwise passes through the four agents in sequence, which purify grooming signals, plan realistic interactions, detect privacy-deceptive behaviors, and repair the interface before the page is cached and rendered. This order embodies our core design insight that the stages reinforce one another along the supply chain: purification hardens all downstream reasoning, the planned interactions decide what the detector sees, and the detected patterns dictate what repair may change, so the pipeline acts as defense in depth rather than a loose collection of detectors. Deployment and offline preparation are detailed in Appendix~\ref{apx:workflow_details}. We now state the four sub-problems and the insight behind each.

\begin{problem}[Grooming purification]\label{prob:purify}
Train a binary provenance classifier $f_{\text{GF}}$ that labels a page as \emph{original} or \emph{generated}, and deploy it as a filter: only pages labeled original are forwarded downstream. A defensive prompt $I_{\text{def}}$ additionally constrains downstream reasoning for pages that evade the filter.
\end{problem}

\noindent The central insight is that the right classification target is \emph{provenance} rather than deceptiveness: a groomed page remains human-benign within the perceptual budget by construction the payload of a groomed page is invisible by construction (Definition~\ref{def:groom}), but the act of generation itself leaves a detectable trace in representation space (Assumption~\ref{ass:provenance}).

\begin{problem}[Budgeted exploration]\label{prob:explore}
Given a site with page set $\mathcal{W}_s$ and a visit budget $B$, generate a task list $T$ whose execution visits pages $W_T\subseteq\mathcal{W}_s$ with $|W_T|\le B$, maximizing the PDP-type coverage
\begin{equation}
\mathrm{Cov}(T)=
\frac{\big|\{y\in\mathcal{Y}: y\in\phi(w) \text{ for some } w\in W_T\}\big|}
     {\big|\{y\in\mathcal{Y}: y\in\phi(w) \text{ for some } w\in\mathcal{W}_s\}\big|}.
\label{eq:pdp_coverage}
\end{equation}
\end{problem}

\noindent The intuition is that PDPs are not spread uniformly over a site. They concentrate along realistic interaction paths such as consent flows, sign-ups, and configuration dialogs, precisely because these are the places where users make privacy-relevant choices. Under a limited budget, coverage is therefore best served by imitating a user rather than crawling exhaustively, which reframes exploration as generating plausible interaction tasks instead of enumerating links.

\begin{problem}[PDP detection]\label{prob:detect}
Learn a detector $\Phi$ that maps a screenshot $s_w$, an expert-validated knowledge base $\mathcal{K}$, and the defensive prompt $I_{\text{def}}$ to a predicted pattern set $\hat{\phi}(w)\subseteq\mathcal{Y}$, maximizing multi-label F1 against the ground truth $\phi(w)$.
\end{problem}

\noindent Framing detection as reasoning over the context $(\mathcal{K},I_{\text{def}})$, rather than as pattern matching, is what makes the detector adaptive to unseen and contextual deceptions. The knowledge base $\mathcal{K}$ injects human heuristics that complement MLLM blind spots, since humans and models are vulnerable to different manipulations~\cite{chen2025obvious,tang2025darkpatternsmeetgui}.

\begin{problem}[Functionality-preserving repair]\label{prob:repair}
Let $\mathcal{A}(y)$ denote the admissible repair actions for PDP pattern $y$, drawn from \{Warning, Highlighting, Block Content, Auto Cookie Setting, UI Cleaning\} as specified in Table~\ref{tab:dp_remidiation}.  Given a webpage $w$ and the predicted PDP set $\hat{\phi}(w)$, produce a repaired webpage~$w^\star$ such that
\begin{align}
H(w^\star) &< H(w) \qquad \text{(harm reduction)} \label{eq:harm}\\
\mathcal{F}(w^\star)&=\mathcal{F}(w) \qquad\text{(functionality preservation)} \label{eq:func}\\
a_y \in \mathcal{A}(y) \ &\ \forall y\in\hat{\phi}(w), \qquad\text{(regulatory admissibility)} \label{eq:legal}
\end{align}
where $H(\cdot)$ denotes the aggregate privacy risk induced by the PDPs present in a webpage, $\mathcal{F}(\cdot)$ denotes task-level functionality, and $a_y$ is the repair action applied to detected pattern~$y$.
\end{problem}

\noindent The three constraints capture repair as a bounded intervention rather than unrestricted rewriting. A repair should reduce PDP-induced privacy risk without disrupting legitimate webpage functionality, while each detected pattern can only be modified using an admissible action from $\mathcal{A}(y)$. This formulation accommodates both awareness-oriented interventions, such as warnings and highlighting, and stronger actions that directly modify deceptive interface elements. The constraints capture the observation that repair is a constrained intervention rather than deletion. Some manipulative elements remain legally legitimate interface components, in which case the strongest admissible response degrades to raising user awareness (Eq.~\eqref{eq:legal}). At the same time, any modification aggressive enough to remove harm can also break the page, which is why the action space must be bounded in advance instead of being left to free-form edits (Eq.~\eqref{eq:func}). Repair is thus a trade-off among harm, functionality, and legality, resolved for each pattern type.

\subsection{Grooming Purifying Agent}
\label{sec:purify}
This agent solves Problem~\ref{prob:purify} with two complementary components: \textit{(i)} a grooming filter $f_{\text{GF}}$ deployed at the input stage to collaborate with the proxy and block suspicious content, and \textit{(ii)} a defensive prompt $I_{\text{def}}$ applied in the PDP Detection Agent (\S\ref{sec:detect}) to prevent grooming samples from misleading downstream reasoning.

\noindent \textbf{Grooming filter.}
Standard pre-processing defenses (\eg denoising or pixel-level comparison) may mitigate basic perturbations such as noise or pixel modification, but fundamentally fail against invisible-text and alpha-channel attacks: by Eq.~\eqref{eq:gap}, the payload of those attacks lives entirely in the semantic gap $\Delta(w)$, so pixel-level inspection reveals no visible differences; the discrepancy emerges only when the model loads the image into memory and reads metadata or alpha channels. This motivates fine-tuning a model to perform provenance classification at the representation level rather than relying on surface-level filtering. Our grooming filter is a supervised classifier fine-tuned from pre-trained vision or multimodal backbones with an additional projection layer that classifies each input as \emph{generated} or \emph{original}. We benchmark several vision backbones, OpenCLIP variants, and commercial MLLMs for provenance classification; EfficientNet-V2-L achieves the highest F1 and is selected as the deployed filter (Appendix Table~\ref{tab:binary_classifier_performance}).

To explain why the filter performs well, we hypothesize that fine-tuning lets the model learn subtle discriminative features embedded in adversarial manipulations, particularly those designed to conceal DPs~\cite{abusnaina2021adversarial}. Although often imperceptible to humans, such features form detectable signatures in representation space~\cite{feinman2017detecting}. To test the hypothesis, we compare the grooming filter with CLIP~\cite{radford2021learning}, a widely used general-purpose encoder, by analyzing how both respond to grooming perturbations. Let $F_{\text{GF}}$ denote the embedding function of our fine-tuned filter, instantiated as an EfficientNet~\cite{tan2019efficientnet} feature extractor, and $F_{\text{CLIP}}$ that of a pre-trained CLIP encoder. For $N$ image pairs, we define the feature-space distances
\begin{equation}
\label{eq:distances}
\begin{aligned}
D_{\text{GF}} &= \left\{ \|F_{\text{GF}}(I_o^{(i)}) - F_{\text{GF}}(I_g^{(i)})\|_2 \right\}_{i=1}^{N},\\
D_{\text{CLIP}} &= \left\{ \|F_{\text{CLIP}}(I_o^{(i)}) - F_{\text{CLIP}}(I_g^{(i)})\|_2 \right\}_{i=1}^{N},
\end{aligned}
\end{equation}
and compute their Pearson correlation
\begin{equation}
\label{eq:pearson}
\rho(D_{\text{GF}}, D_{\text{CLIP}}) = \frac{\text{Cov}(D_{\text{GF}}, D_{\text{CLIP}})}{\sigma_{D_{\text{GF}}} \cdot \sigma_{D_{\text{CLIP}}}}.
\end{equation}
A positive correlation indicates that the two encoders respond consistently to grooming-induced representation shifts. Since CLIP is trained to encode high-level semantic features~\cite{radford2021learning}, strong agreement indicates that the filter captures semantically meaningful, task-relevant signals rather than overfitting to superficial artifacts; the correlation thus provides statistical evidence that the filter learns generalizable representations of grooming manipulations, empirically discharging Assumption~\ref{ass:provenance}. 

Given CLIP's strong performance as a general-purpose visual encoder, it is natural to consider using it directly for grooming detection. To test this baseline, we append a binary classification head~\cite{chen2020simple,hendrycks2016baseline} to the CLIP image encoder. However, as shown in Table~\ref{tab:llm_groom_attack_and_defense}, our fine-tuned grooming filter significantly outperforms this CLIP-based classifier, demonstrating the advantage of task-specific fine-tuning for capturing nuanced grooming signals.

\noindent\textbf{Defensive prompt.}
Because the multimodal LLM (MLLM) used in \tool{} may be a commercial black-box model that cannot be fine-tuned, we further introduce a prompt-based defense. We design a carefully crafted instruction $I_{\text{def}}$ (Appendix~\ref{apx:grooming_defense_prompt}) that explicitly alerts the model to potential grooming scenarios and constrains its behavior during PDP detection: it presents both original and generated examples and instructs the model to reason about subtle manipulations and to prioritize privacy-harm detection under adversarial settings. If the model has already been partially compromised, its outputs may reflect grooming-induced biases; otherwise, $I_{\text{def}}$ serves as a strict reasoning scaffold that guides the LLM to resist misleading inputs and accurately identify PDPs.

\subsection{Task Generation Agent}
\label{sec:taskgen}
To solve Problem~\ref{prob:explore}, we adopt a reinforcement learning (RL) framework that optimizes prompts for generating realistic and diverse exploration tasks: a summarization model first extracts the high-level purpose of the target website to condition prompt generation, and the RL agent iteratively refines instructions to produce task lists that maximize both coverage of known behaviors and novelty of useful variations. Ground-truth task lists are constructed as detailed in Appendix~\ref{apx:dataset_collection}.

\noindent \textbf{Information extraction model.}
Because ``About Us'' pages vary widely in style and availability, we query the Google search engine with the site URL via Serper~\cite{serper2025}; from the top 10 results (titles and snippets), a summarization LLM (GPT-4o) synthesizes a concise description of the website's purpose, which is provided as contextual input to the task generation pipeline so that generated tasks align with the site's real-world use case.

\noindent \textbf{Multi-agent RL-based task generation.}
Inspired by PrivAgent~\cite{nie2024privagent}, which introduced a reinforcement learning framework to extract system prompts and training data from target LLMs, we adapt and extend this idea to the task generation setting. Our objective differs: rather than extracting internal information, we aim to generate high-quality exploration tasks by learning optimized instruction prompts through Proximal Policy Optimization (PPO)~\cite{schulman2017ppo}, a stable policy-based method that directly optimizes the policy using clipped surrogate objectives. This component integrates five tightly coupled elements, and the full procedure is summarized in Algorithm~\ref{alg:rl}.

\begin{algorithm}[t]
\caption{RL-based task generation prompt optimization.}
\label{alg:rl}
\small
\KwIn{Initial prompt $I_0$; training dataset~$\mathcal{D}=\{(x_i,T_i)\}_{i=1}^{N}$; pre-trained policy model $\pi_\theta$; target multimodal model $M$; task-matching threshold $\tau_t$; prompt-similarity threshold $\tau_s$; reward threshold $\tau_r$; reward weight $p$; bonus reward $\alpha$; total epochs $E$.}
\KwOut{Optimized prompt $I^\star$.}

Initialize PPO agent $\mathrm{TRL}\leftarrow\mathrm{PPO}(\pi_\theta)$\;
Initialize high-reward prompt set $P\leftarrow\emptyset$\;
Initialize $I^\star\leftarrow I_0$ and $r^\star\leftarrow-\infty$\;

\For{$\mathrm{epoch}=1$ \KwTo $E$}{
    \ForEach{batch $(X,T_G)\subset\mathcal{D}$}{
        Generate candidate prompt $I_{\mathrm{cand}}\sim\mathrm{TRL}(I_0)$\;
        Predict tasks $T_P\leftarrow M(X,I_{\mathrm{cand}})$\;
        Compute reward $r$ using Eqs.~(\ref{eq:sim})--(\ref{eq:reward})\;

        \If{$P\neq\emptyset$ \textbf{and} $\mathrm{SIM}_p(I_{\mathrm{cand}},P)>\tau_s$}{
            $r\leftarrow r+\alpha$\;
        }

        \If{$r>\tau_r$}{
            $P\leftarrow P\cup\{I_{\mathrm{cand}}\}$\;
        }

        \If{$r>r^\star$}{
            $I^\star\leftarrow I_{\mathrm{cand}}$\;
            $r^\star\leftarrow r$\;
        }

        $\mathrm{TRL.step}(I_0,I_{\mathrm{cand}},r)$\;
    }
}
\Return $I^\star$\;
\end{algorithm}

\begin{itemize}[leftmargin=*]
\item \textit{LLM RL agent ($T_{\text{RL}} = \mathrm{PPO}(\pi_\theta)$).} The agent generates high-quality textual instructions that guide the target MLLM in producing exploration tasks. We adopt Qwen QwQ-32B~\cite{qwq32b} as the policy network $\pi_\theta$, which offers strong reasoning capabilities at a manageable size of 32 billion parameters, making it suitable for efficient RL training, and follow the PPO setup introduced in PrivAgent~\cite{nie2024privagent}. Training begins with an initial instruction prompt $I_0$, refined through iterative interaction with the environment to produce candidate prompts $I_\text{cand}$ that maximize the reward.
\item \textit{Target MLLM ($M$).} Each candidate prompt $I_\text{cand}$ is sent to the target multimodal LLM together with two additional inputs: a screenshot of the current webpage and the summarized website purpose from the information extraction model. We select OpenAI o4-mini as the target MLLM for its multimodal input support and strong reasoning performance; it outputs the predicted task set $T_P$ used to evaluate the effectiveness of the current prompt.
\item \textit{Similarity calculation.} For a ground-truth task $t$ and predicted task $t'$, we normalize task-specific entities using $\textsc{EP}(\cdot)$ and compute a modified Levenshtein Token Set Ratio (TSR). A pair is considered matched when its similarity exceeds threshold $\tau_t$:
\begin{equation}
\begin{aligned}
\mathrm{SIM}(t,t') &= \textsc{TSR}\!\left(\textsc{EP}(t),t'\right),\\
m(t,t') &= \mathbf{1}\!\left[\mathrm{SIM}(t,t')\geq\tau_t\right].
\end{aligned}
\label{eq:sim}
\end{equation}

\item \textit{Reward function.} We measure \emph{coverage} as the fraction of ground-truth tasks matched by at least one predicted task, and \emph{novelty} as the fraction of predicted tasks unmatched by the ground-truth set:
\begin{equation}
\begin{aligned}
\mathrm{cov}(T_G,T_P) &= \frac{1}{|T_G|}\sum_{t\in T_G}\max_{t'\in T_P}m(t,t'),\\
\mathrm{nov}(T_G,T_P) &= 1-\frac{1}{|T_P|}\sum_{t'\in T_P}\max_{t\in T_G}m(t,t').
\end{aligned}
\label{eq:coverage_novelty}
\end{equation}
The reward is then defined as
\begin{equation}
r = p\,\mathrm{cov}(T_G,T_P) + (1-p)\,\mathrm{nov}(T_G,T_P),
\label{eq:reward}
\end{equation}
where $p\in[0,1]$ balances coverage and novelty, incentivizing the RL agent not only to replicate known tasks but also to discover novel yet relevant ones.
\item \textit{Bonus reward ($\alpha$).} To encourage reuse of effective prompting patterns, we maintain a set $P$ of high-reward prompts whose reward exceeds $\tau_r$. A candidate prompt receives a bonus if it is sufficiently similar to any prompt in $P$:
\begin{equation}
\label{eq:bonus}
r \leftarrow
\begin{cases}
r+\alpha, & \mathrm{SIM}_p(I_{\mathrm{cand}},P)>\tau_s,\\
r, & \text{otherwise},
\end{cases}
\end{equation}
where $\mathrm{SIM}_p(I_1,I_2)=\textsc{TSR}(I_1,I_2)$ and $\mathrm{SIM}_p(I_{\mathrm{cand}},P)=\max_{I\in P}\mathrm{SIM}_p(I_{\mathrm{cand}},I)$. Here, $\tau_s$ is the prompt-similarity threshold.
\end{itemize}

\subsection{PDP Detection Agent}\label{sec:detection_agent}
\label{sec:detect}
This agent realizes the detector $\Phi(s_w,\mathcal{K},I_{\text{def}})$ of Problem~\ref{prob:detect} by combining an automated browsing agent that simulates real user interactions, an expert-validated knowledge base~$\mathcal{K}$, and a reasoning-capable MLLM. The browsing agent executes the task list from the Task Generation Agent; each time a screenshot action is triggered, the screenshot is sent both back to the automation model and to the detection model to determine whether any PDPs are present.

\noindent \textbf{Website automation model.} Existing policy-based or environment-specific agents~\cite{he2024webvoyager,zhang2025webpilot,qin2025ui,wu2023webui,sherin2023qexplore} are either ineffective at simulating daily user activity or restricted in the resources they can access. For extensibility, \tool{} employs a computer-use agent built upon Anthropic's Computer Use demo~\cite{anthropic_computer_use}, translating natural-language task lists into concrete actions on live websites, designed as a plug-and-play module.

\noindent \textbf{Definition generation agent.} This module builds the knowledge base $\mathcal{K}$: structured, domain-aware descriptions of deceptive patterns extracted with LLMs from public taxonomies~\cite{ChatGPTLearning}, regulatory guidelines, and developer documentation. In DPGuard~\cite{shi202550}, the prompts describing each pattern are derived from its taxonomy alone. To address this, our approach {integrates human heuristics with LLM-based reasoning} through a two-stage expert annotation pipeline: five domain experts, each with over five years of experience in AI security and privacy, independently annotated deceptive patterns using our PDP taxonomy (average pairwise Cohen's $\kappa=0.85$), with disagreements resolved by strict majority voting; the experts then systematically sampled the taxonomy to curate concrete, real-world AI grooming examples for each category. This expert-validated dataset, combining consensus interpretations, deployment contexts, and empirical samples, constitutes $\mathcal{K}$ and, together with $I_{\text{def}}$, strictly steers the LLM's downstream reasoning.

\noindent \textbf{MLLMs for PDP detection with reasoning.} 
We use Gemini 2.5 Pro~\cite{Kavukcuoglu2025Gemini} as the reasoning-capable MLLM for PDP detection, selected based on its strong performance in LM Arena~\cite{lmarena}.

\subsection{Interface Repairing Agent}
\label{sec:repair}

The Interface Repairing Agent realizes Problem~\ref{prob:repair} by translating each detected PDP into an admissible modification of the corresponding webpage. Given the predicted pattern set~$\hat{\phi}(w)$, the agent first identifies the deceptive UI elements associated with each detected pattern and then selects a repair action from $\mathcal{A}(y)$. The repair space is restricted to five predefined actions, \ie \textit{Warning}, \textit{Highlighting Important Links}, \textit{Block Content}, \textit{Auto Cookie Setting}, and \textit{UI Cleaning}, thereby directly operationalizing the functionality-preservation and regulatory-admissibility constraints in Eqs.~\eqref{eq:func} and~\eqref{eq:legal}.

The design of these actions is grounded in Protection Motivation Theory (PMT)~\cite{rogers1975protection}, which has been widely applied to security and privacy interventions~\cite{van2019using,prange2022secure,chaudhary2024driving,lu2024awareness}. In particular, the repair actions span two complementary forms of protection. \emph{Awareness-oriented} interventions, such as warnings and highlighting privacy-critical links, preserve user autonomy while making deceptive or obscured information more salient. \emph{Action-oriented} interventions directly reduce privacy harm when stronger modification is appropriate, for example by blocking undisclosed advertisements, restoring cookie choices to the user's expressed preference, or removing visual asymmetries that bias users toward a preferred option. This distinction allows \tool{} to mitigate deceptive influence without indiscriminately removing legitimate interface functionality.

Table~\ref{tab:dp_remidiation} specifies the admissible repair actions for each PDP type. A computer-use model applies the selected action to the identified interface elements while preserving the surrounding webpage structure. Detailed proxy, caching, and repair procedures are provided in Appendix~\ref{apx:repair_actions}.

\begin{table}[t]
\centering
\caption{Repair approaches.}
\label{tab:dp_remidiation}
\resizebox{\linewidth}{!}{
\begin{tabular}{llccccc}
\toprule
\multirow{4}{*}{\textbf{Privacy Types}} & 
\multirow{4}{*}{\textbf{Deceptive Patterns}} &
\multicolumn{5}{c}{\textbf{Repair Approach}}\\
\cmidrule(){3-7}
& & \textbf{Warning} &
\makecell{\textbf{Highlighting}\\\textbf{Important}\\\textbf{Links}} & 
\makecell{\textbf{Block}\\\textbf{Content}} & 
\makecell{\textbf{Auto}\\\textbf{Cookie}\\\textbf{Setting}} & 
\makecell{\textbf{UI}\\\textbf{Cleaning}} \\
\midrule
\multirow{3}{*}{\begin{tabular}{@{}l@{}}Information\\Concealment\end{tabular}} 
 & Hidden Information  &\pie{360} &\pie{360} &\pie{0}&\pie{0}&\pie{0}\\
 & Disguised Ads - wo Hint&\pie{0} &\pie{0} &\pie{360}&\pie{0}&\pie{0}\\
 & Disguised Ads - w Hint &\pie{360} &\pie{0}&\pie{0}&\pie{0}&\pie{0}\\
\midrule
\multirow{3}{*}{\begin{tabular}{@{}l@{}}Interface\\Obstruction\end{tabular}} 
 & Pre-Selection&\pie{360}& \pie{0} &\pie{0}&\pie{0}& \pie{0}\\
 & False Hierarchy&\pie{0} &\pie{0}&\pie{0}& \pie{360}& \pie{360}\\
 & Small Close Button  & \pie{360} &\pie{0}&\pie{0}&\pie{0}&\pie{0}\\
\midrule
\multirow{2}{*}{\begin{tabular}{@{}l@{}}Forced Data\\Disclosure\end{tabular}} 
 & Privacy Zuckering& \pie{360} &\pie{360} &\pie{0}& \pie{0}&\pie{0}\\
 & Forced Action& \pie{360}&\pie{360}&\pie{0}&\pie{0}&\pie{0}\\
\bottomrule
\end{tabular}
}
\end{table}

%===============================================================================
\section{Experimental Evaluation}
\label{sec:eval}
In this section, we evaluate defense against AI grooming, task list generation, website exploration, PDP detection, and interface repair, followed by an empirical study in the wild. As laid out in Section~\ref{sec:problem}, each subsection reports the metric induced by its corresponding formal problem.

\subsection{Experimental Setup}
\label{sec:setup}

\noindent\textbf{Datasets.}
We construct four datasets for training and evaluating \tool{}:
\textit{(1) Task Generation Dataset.} We derive website-task pairs from Multimodal-Mind2Web~\cite{deng2023mindweb} and, after consolidating tasks associated with similar screenshots, obtain 112 training and 40 testing pairs for task generation and website exploration;
\textit{(2) PDP Detection Dataset.} Starting from the 40 test websites, \tool{} executes 258 generated tasks and collects 418 webpage screenshots. The same five experts of Section~\ref{sec:detection_agent} annotate them under our PDP taxonomy, yielding 233 benign webpages and 185 webpages (282 instances) with PDPs, containing 515 labeled instances in total;
\textit{(3) Grooming Dataset.} We construct grooming samples under the two attack settings of Section~\ref{sec:attacks}. For the \emph{naive whole-page baseline}, GPT-based image editing generates visually similar counterparts of the PDP Detection Dataset, yielding 233 generated benign and 182 generated PDP screenshots. For the \emph{targeted grooming attacks}, we identify 33 high-risk privacy-deceptive zones and apply the five grooming strategies, yielding 8,181 samples; and
\textit{(4) Wild Dataset.} For the empirical study in the wild (\S\ref{sec:eval_wild}), we collect 485 unique top-ranked websites spanning 14 categories.
Full collection and generation details are provided in Appendices~\ref{apx:dataset_collection} and~\ref{apx:grooming_dataset}.

\noindent\textbf{Baselines and metrics.} For exploration, we compare against Breadth-First Search, random exploration~\cite{stafeevSoKState25}, and the policy-based WebUI~\cite{wu2023webui}, measuring PDP-type coverage (Eq.~\eqref{eq:pdp_coverage}) and pages visited. For detection, we compare against DPGuard~\cite{shi202550}, ALSACNC~\cite{alsacnc}, and AidUI~\cite{mansur2023aidui}, reporting micro- and macro-averaged F1 (Appendix~\ref{apx:benchmarks}). 
For targeted grooming, we report the PDP detection rate, defined as the fraction of manipulated samples for which the downstream detector still correctly identifies the underlying PDP; higher is better for the defender.

\noindent\textbf{Models.} The grooming filter is a fine-tuned EfficientNet-V2-L; the RL policy is Qwen QwQ-32B with OpenAI o4-mini as the target MLLM and GPT-4o for website summarization; detection uses Gemini 2.5 Pro; and UI automation and repair use Claude 3.7 Sonnet Computer Use. Runtime hardware and per-page latencies are reported in Appendix~\ref{apx:grooming_filter_performance}.

\subsection{Defense Against Grooming Samples}
\label{sec:eval_grooming}

\begin{table}[t]
\centering
\caption{PDP detection performance against grooming attack.}
\label{tab:llm_groom_on_exploration_dataset}
\resizebox{\linewidth}{!}{
\begin{threeparttable}
\begin{tabular}{@{}lccccccccc@{}}
\toprule
\multirow{2}{*}{} & \multirow{2}{*}{Dataset} & \multirow{2}{*}{\begin{tabular}[c]{@{}c@{}}\# of\\Samples\end{tabular}} & \multirow{2}{*}{\begin{tabular}[c]{@{}l@{}}Filter \\ Rate\end{tabular}} & \multicolumn{3}{c}{w/o Defense Prompt} & \multicolumn{3}{c}{w/ Defense Prompt} \\
\cmidrule(l){5-7}\cmidrule(l){8-10}
 &  &  &  & Precision & Recall & F1 & Precision & Recall & F1 \\
\midrule
\multirow{2}{*}{w/o GF} & Original & 233 & - & 0.9406 & 0.8155 & 0.8736 & - & - & - \\
 & Generated & 233 & - & 1.0000 & 0.4217 & 0.5933 & 1.0000 & 0.4826 & 0.6510 \\
\midrule
w/ GF & Generated & 21/233 & 90.98\% & 1.0000 & 0.2381 & 0.3846 & 1.0000 & 0.3333 & 0.5000 \\
\bottomrule
\end{tabular}
\begin{tablenotes}
\item[]GF: Grooming filter; Original: non-AI-generated images.
\end{tablenotes}
\end{threeparttable}
}
\end{table}

We evaluate whether the Grooming Purifying Agent can identify and filter suspicious samples at the input boundary, and whether defensive prompting can mitigate residual samples that evade filtering, using PDP detection as a representative downstream task to quantify the influence of grooming samples.

\noindent\textbf{Downstream impact and sample filtering.}
We first verify that the constructed grooming samples can materially affect an AI consumer. On the \emph{naive whole-page generated sample subset} of the Grooming Dataset (Appendix~\ref{apx:grooming_dataset}), the PDP detector achieves an F1 score of 0.8736 on 233 original benign screenshots, which decreases to 0.5933 on their generated counterparts, a 28.03\% drop. This controlled case study shows that grooming samples can introduce machine-consumable signals that alter downstream reasoning, without implying the same degradation across arbitrary AI tasks. On the same naive generated sample subset, the fine-tuned EfficientNet~\cite{tan2019efficientnet} provenance classifier achieves 0.9022 micro- and macro-F1 and blocks 90.98\% ($=\frac{233-21}{233}$) of generated benign samples before downstream processing.

\noindent\textbf{Defense in depth.}
The correlation analysis of Eq.~\eqref{eq:pearson} yields consistently positive correlations across grooming strategies (Table~\ref{tab:correlation} in Appendix), supporting Assumption~\ref{ass:provenance}. Because filtering is not perfect, \tool{} additionally applies the defensive prompt $I_{\text{def}}$ to residual samples. In the PDP detection case study, defensive prompting improves F1 on groomed samples by 0.06, while combining filtering and prompting yields a further 0.12 improvement over filtering alone.

\begin{table}[t]
\centering
\caption{PDP detection rates against targeted grooming.}
\label{tab:llm_groom_attack_and_defense}
\resizebox{\linewidth}{!}{
\begin{tabular}{@{}cccccc@{}}
\toprule
\makecell{\textbf{Defense}\\\textbf{Type}} & 
\makecell{\textbf{Noise}\\\textbf{Generation}} & 
\makecell{\textbf{Pixel}\\\textbf{Modification}} & 
\makecell{\textbf{Invisible}\\\textbf{Text}} & 
\makecell{\textbf{Alpha}\\\textbf{Channel}} & 
\makecell{\textbf{Unreadable}\\\textbf{Text}} \\
\midrule
% \multicolumn{6}{c}{Attack Success Rate} \\
% \midrule
No Grooming Defense & 0.5152 & 0.5273 & 0.4728 & 0.2061 & 0.4788\\
\makecell{CLIP Detection\\(openai/clip-vit-base-patch16)} & 0.6061 & 0.6122 & 0.5394 & 0.3637 & 0.5091 \\
\midrule
Grooming Filter & 0.6425 & 0.6182 & 0.6243 & 0.3819 & 0.5637\\
Defense Prompt & 0.5213 & 0.5394 & 0.4909 & 0.2364 & 0.4848\\
Grooming Filter + Defense Prompt & 0.6606 & 0.6364 & 0.6304 & 0.4061 & 0.5697\\
\bottomrule
\end{tabular}
}
\end{table}

\noindent\textbf{Robustness across grooming strategies.}
We finally evaluate the two defenses under the substantially stronger \emph{targeted grooming attacks}, where the five localized manipulation strategies of Section~\ref{sec:attacks} yield lower detection rates and therefore provide a more challenging robustness setting. As shown in Table~\ref{tab:llm_groom_attack_and_defense}, filtering and defensive prompting are individually beneficial, while their combination consistently provides the strongest protection.

\begin{takeaway}\addtocounter{takeaway}{1}
\textbf{Takeaway \thetakeaway:~}
\textit{\tool{} mitigates grooming at the input boundary rather than modifying the downstream model. The provenance filter blocks 90.98\% of naive whole-page generated samples, while defensive prompting provides complementary protection for residual samples. Under the stronger targeted grooming attacks, combining both mechanisms consistently achieves the strongest defense.}
\end{takeaway}

\subsection{RL Training in Task Generation Agent}

During RL-based prompt optimization, we evaluate model performance on the test set at the end of each batch; training consists of 50 batches over 5 epochs, with 10 batches per epoch. Figure~\ref{fig:batch_reward} in Appendix~\ref{apx:rl_details} presents the reward (Eq.~\eqref{eq:reward}) per batch and its epoch-level average, showing a clear upward trend with both metrics reaching their highest values by the final epoch.

The average rewards are presented in Table~\ref{tab:rf_model_eval} in Appendix~\ref{apx:rl_details}. The $2{\times}2$ ablation study demonstrates that both components, website information extraction and RL-based prompt optimization, contribute independently and synergistically to task generation quality. Disabling both modules yields the lowest reward of 0.3350. Enabling only the RL-engineered prompt nearly doubles this baseline to 0.6626, a 97.9\% increase; enabling only the web-level context raises the reward to 0.4585 (+36.8\%). Critically, enabling both increases the reward to 0.7140, outperforming the prompt-only configuration by 7.8\% and the web-info-only configuration by 55.8\%. These results confirm that integrating website context with an optimized prompt is essential for maximizing \tool{}'s effectiveness on task generation.

\begin{takeaway}\addtocounter{takeaway}{1}
\textbf{Takeaway \thetakeaway:~}
\textit{To assist MLLMs in generating more reliable and executable tasks, both website information and RL-based prompt optimization are essential.}
\end{takeaway}

\subsection{Website Exploration Efficiency}
\label{sec:eval_explore}
\begin{table}[t]
\centering
\caption{Performance of exploration approaches.}
\label{tab:exp_app_cmp}
\resizebox{.9\linewidth}{!}{
\begin{tabular}{lcc}
\toprule
\textbf{Approach} & \textbf{\# of Visited Webpages} & \textbf{Coverage of DP Instances}\\
\midrule
BFS & 4395 & 0.6667 {\small$({=}\,86/129)$}\\
Random Exploration~\cite{stafeevSoKState25} & 2797 & 0.4729 {\small$({=}\,61/129)$}\\
WebUI~\cite{wu2023webui} & 3420 & 0.3721 {\small$({=}\,48/129)$}\\
\textbf{DPAgent (Ours)} & 418 & \textbf{0.8062} {\small$({=}\,104/129)$}\\
\bottomrule
\end{tabular}
}
\end{table}
To evaluate exploration efficiency, we compare \tool{} with the brute-force Breadth-First Search (BFS), random exploration~\cite{stafeevSoKState25}, and the policy-based WebUI~\cite{wu2023webui}; performance is summarized in Table~\ref{tab:exp_app_cmp}. All four approaches were run on the same test dataset of 40 hostnames. For each hostname, we manually recorded the presence of each of the seven predefined PDP types; after merging the results of all strategies, we identified 129 PDP-type instances out of a possible 280 (7 types $\times$ 40 hostnames).

We define PDP-type coverage as the number of PDP-type instances detected by an approach divided by the 129 instances observed across all methods. For time efficiency, we limited BFS and random exploration to a maximum depth of 3 and capped the exploration list at 200 URLs per domain for random exploration and WebUI. Under these constraints, BFS achieved 66.67\% coverage over 4,395 webpages, random exploration 47.29\% over 2,797, and WebUI 37.21\% over 3,420. In contrast, \tool{} achieved the highest coverage, 80.62\%, while exploring only 9.5\%, 14.9\%, and 12.2\% of the pages visited by BFS, random exploration, and WebUI, respectively. These results demonstrate that \tool{}'s dynamic, user-simulated exploration is more effective than brute-force and policy-based methods under time and resource constraints.

\begin{takeaway}\addtocounter{takeaway}{1}
\textbf{Takeaway \thetakeaway:~}
\textit{\textup{\tool{}} adopts a user-activity-simulated website exploration approach and achieves a PDP-type coverage rate of 80.62\%, which is 13.95\%, 33.33\%, and 43.41\% higher than BFS, random exploration, and WebUI, while visiting only 9.5\%, 14.9\%, and 12.2\% as many webpages, respectively.}
\end{takeaway}

\subsection{PDP Detection}
\label{sec:eval_detect}
We compare \tool{} with 3 state-of-the-art PDP detectors using the PDP detection dataset which contains 515 instances spanning benign and deceptive types. For each model, we excluded any pattern category it does not support and used the F1 score, which accounts for both false positives and false negatives. The results are presented in Table~\ref{tab:dp_detection_performance}: \tool{} achieves 
achieves the highest reported micro- and macro-F1 among the evaluated systems under their supported categories.
a new state-of-the-art performance with a micro-average F1 of 0.8162 and a macro-average F1 of 0.6914, improvements of 22.72\% and 26.72\% over the current SOTA model, DPGuard. Across all specific PDP categories, \tool{} falls short only on the Privacy Zuckering category.

\begin{takeaway}\addtocounter{takeaway}{1}
\textbf{Takeaway \thetakeaway:~}
\textit{\textup{\tool{}} employs a complementary definition generation strategy that integrates grooming samples and human expertise with LLM-based agent understanding and enhanced reasoning. This significantly boosts detection performance, setting a new SOTA with a micro-average F1 of 0.8162 and a macro-average F1 of 0.6914.}
\end{takeaway}

\begin{table}[t]
\centering
\caption{Performance of PDP detection.}
\label{tab:dp_detection_performance}
\resizebox{\linewidth}{!}{
\begin{tabular}{llccccc}
\toprule
\textbf{Type} & 
\textbf{Deceptive Patterns} & 
\textbf{Instances} & 
\makecell{\textbf{DPGuard}\\\textbf{\cite{shi202550}}} & 
\makecell{\textbf{AidUI}\\\textbf{\cite{mansur2023aidui}}} & 
\makecell{\textbf{ALSACNC}\\\textbf{\cite{alsacnc}}} & 
\makecell{\textbf{DPAgent}\\\textbf{(Ours)}} \\
\midrule
No DP & No DP& 233 & 0.7140 & 0.6150 & 0.5181 & \textbf{{0.8736}} \\
\midrule
\multirow{2}{*}{\shortstack[l]{Information\\Concealment}} 
 & Hidden Information & 63 & 0.2105 & - & 0.2035 & \textbf{{0.8143}} \\
 & Disguised Ads & 106 & 0.7273 & 0.2837 & - &\textbf{{0.9223}} \\
\midrule
\multirow{3}{*}{\shortstack[l]{UI\\Manipulation}} 
 & Preselection & 8 & 0.2381 & 0.0635 & - & \textbf{{0.5714}} \\
 & False Hierarchy & 25 & 0.2759 & 0.0000 & - & \textbf{{0.6364}} \\
 & Small Close Button & 28 & 0.3590 & - & - & \textbf{{0.7111}} \\
\midrule
\multirow{2}{*}{\shortstack[l]{Forced\\Data Disclosure}}
 & Privacy Zuckering & 6 & \textbf{{0.4444}} & 0.0000 & 0.2222 & 0.2857 \\
 & Forced Action & 46 & - & - & 0.1538 & \textbf{{0.7167}} \\
\midrule
\midrule
\multicolumn{2}{c}{micro avg} &\multirow{2}{*}{515} & 0.5890 & 0.4315 & 0.3449 & \textbf{{0.8162}} \\
\multicolumn{2}{c}{macro avg} && 0.4242 & 0.1924 & 0.2744 & \textbf{{0.6914}} \\
\bottomrule
\end{tabular}}
\end{table}

\subsection{Robustness Against Adaptive Attackers}
\label{sec:eval_adaptive}

The 90.98\% filtering rate in Section~\ref{sec:eval_grooming} considers naive whole-page generated samples. We next evaluate the stronger gray-box adaptive setting defined in Definition~\ref{def:adaptive}.

\noindent\textbf{Gray-box transfer.}
We measure robustness using two metrics: the \emph{filter bypass rate}, the fraction of groomed samples classified as \emph{original} by $f_{\text{GF}}$, and the \emph{attack success rate} (ASR), the fraction that both bypass the filter and continue to mislead the downstream PDP detector. Using ResNet-152, ConvNeXt, and OpenCLIP as surrogate encoders, transferred perturbations achieve a bypass rate of only 19.6\%, while the ASR is 8.4\% across the five grooming strategies in Section~\ref{sec:attacks}.

\noindent\textbf{Residual protection.}
Applying the defensive prompt $I_{\text{def}}$ to samples that bypass the filter further reduces ASR by 0.8 percentage points, corresponding to a 9.5\% relative reduction. This provides an additional layer of protection against residual adaptive evasions.

\begin{takeaway}\addtocounter{takeaway}{1}
\textbf{Takeaway \thetakeaway:~}
\textit{Gray-box adaptive attacks transfer poorly to the deployed filter: only 19.6\% bypass the filter, and just 8.4\% retain downstream attack effectiveness. The defensive prompt further reduces the residual ASR, providing layered protection against adaptive evasion.}
\end{takeaway}

\subsection{Automatic Interface Repair}
\label{sec:eval_repair}

After PDPs are captured, automatic interface repair is triggered (Problem~\ref{prob:repair}). We first measure the repair rate across the 418 webpages of the detection dataset; to further assess effectiveness and impact on the browsing experience, we conducted a user study with eight use cases and 40 global participants.

\noindent\textbf{Repair rate.}
Among the 418 webpages, there are 233 benign and 282 deceptive instances. \tool{} correctly identified 243 of the 282 deceptive instances (86.17\%). We then applied the repair actions of Section~\ref{sec:repair} and manually reviewed the 243 detected instances; this manual review is how the functionality-preservation constraint~\eqref{eq:func} is verified in practice. 
Of the 243 correctly detected instances, 218 were successfully repaired, corresponding to a conditional repair success rate of 89.71\% and an end-to-end repair rate of 77.30\% over all 282 PDP instances, where success means an admissible action satisfying constraints \eqref{eq:harm}-\eqref{eq:legal} was found and applied. Among the failures, Disguised Ads without hints were the most difficult to address, for two reasons: the ambiguous decision boundary when specifying the advertisement's location, and complex HTML element hierarchies that prevent removing the element from pixel-level information alone. These are failures of localization rather than of the repair action set.

\begin{figure}[t]
    \centering
    \includegraphics[width=1.0\linewidth]{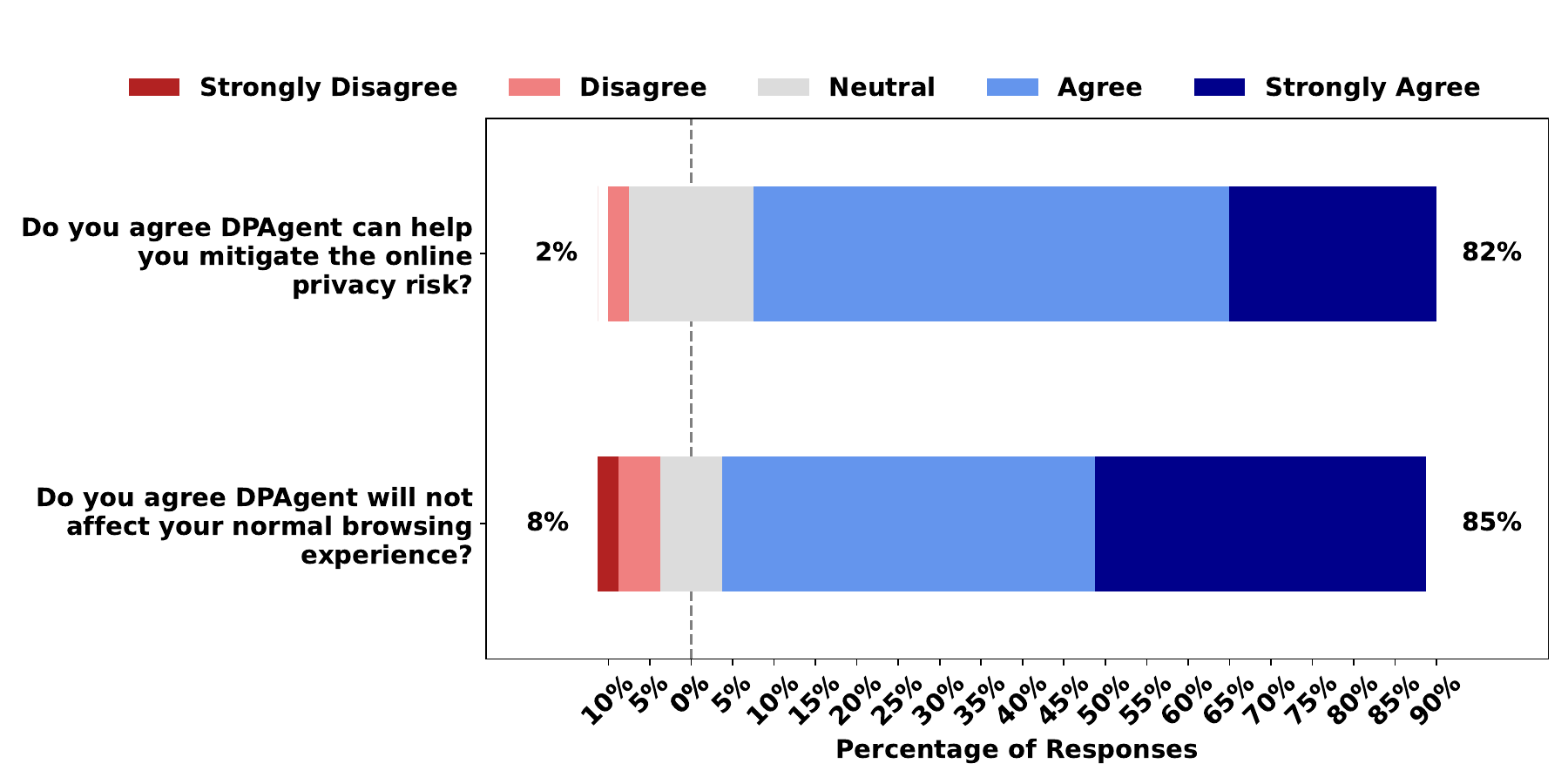}
    \caption{User study evaluating the effectiveness of interface repair and its impact on the browsing experience.}
    \label{fig:user_study}
\end{figure}

\noindent\textbf{User study.}
We conducted a user study assessing both effectiveness and impact on the browsing experience, rated from 1 (strongly negative) to 5 (strongly positive). Figure~\ref{fig:user_study} shows the rating distributions across 40 participants: the average effectiveness rating was 4.10 (55\% rated above 4), and the average rating for impact on the browsing experience was 4.16 (60\% rated above 4). Both scores suggest that \tool{} mitigates PDP risks while enhancing the overall browsing experience. Study protocol details, including participants' privacy concern levels and prior exposure to PDPs, are provided in Appendix~\ref{apx:user_study}; real-world before-and-after repair examples appear in Appendix~\ref{apx:before_after_repair}.

\begin{takeaway}\addtocounter{takeaway}{1}
    \textbf{Takeaway \thetakeaway:~}
    \textit{\textup{\tool{}} achieved an end-to-end repair rate of 0.7730 on the privacy deceptive pattern detection dataset. The average effectiveness rating was 4.10 and the average browsing-experience rating was 4.16, indicating that participants perceived the repairs as effective while causing limited disruption to normal browsing.}
\end{takeaway}

\subsection{Empirical Evaluation in the Wild}
\label{sec:eval_wild}
We evaluate DPAgent in the wild across website-level, pattern-level, and geographic dimensions.

\noindent\textbf{Website-level study.}
We ran \tool{} on 485 selected websites from the wild dataset and obtained 5,206 webpages. Browsing only homepages, 167 websites (34.43\%) contained PDPs; manually checking 17 randomly selected sites from this set validated that 94.11\% do contain PDPs. When dynamically exploring the entire websites, 476 (98.14\%) were found to include at least one PDP. Attempting homepage repairs, most could be addressed, but 47 websites (9.69\%) could not be successfully repaired, primarily due to complex HTML element hierarchies.

\noindent\textbf{Pattern-level study.}
Our analysis revealed that the top three most frequently violated PDPs are Hidden Information (18.2\%), Forced Action (15.6\%), and Disguised Ads (10.6\%). Defining a PDP as repairable only if it can be detected, Disguised Ads emerged as the most difficult type to repair.

\noindent\textbf{Geographic study.}
To examine how PDP risk varies across regions, we filtered the top 50 websites by domain suffixes for Asia, America, Europe, and Oceania, and analyzed homepage PDPs only. Prevalence is: Oceania (84\%), Asia (80\%), America (78\%), and Europe (70\%). 

\begin{takeaway}\addtocounter{takeaway}{1}
    \textbf{Takeaway \thetakeaway:~}
    \textit{The empirical study reveals that 34.43\% of websites contain at least one privacy deceptive pattern on their homepage; with thorough exploration this rate increases to 98.14\%. Applying \textup{\tool{}}, more than 90\% of these PDPs can be repaired. Hidden Information is the most frequently violated pattern, Disguised Ads the most difficult to repair, and Oceania exhibited the highest observed homepage PDP rate in our sample.}
\end{takeaway}

%===============================================================================

\section{Related Work}\label{sec:related}
We summarize the most relevant work below, with additional details in Appendix~~\ref{apx:related} and a comparison in Table~\ref{tab:related_work}.

\noindent \textbf{Data voids and AI grooming.}
Prior studies examine how data voids~\cite{boyd2019data} and coordinated content campaigns can influence search, retrieval, and LLM outputs~\cite{newsguard2025pravda,radina2025grooming,prorussia2025american}, while a recent Dagstuhl working group highlights how generative AI may further scale such influence operations~\cite{guennemann2026securityprivacyllm}. Alyukov~\etal, however, find limited evidence of deliberate manipulation and attribute chatbot references to Kremlin-linked sources mainly to information gaps~\cite{alyukov2025llms}. Our work distinguishes organic scarcity from adversarial manipulation (Definitions~\ref{def:scg} and~\ref{def:ufg}), links data void exploitation to PDPs, and defends the point where groomed artifacts enter the data pipeline.

\noindent \textbf{Website privacy and deceptive pattern detection.}
Prior tools target network level tracking and intrusions~\cite{siby2019encrypted,dahlberg2023timeless,singh2024connecting,xue2025theDis}, as well as UI threats such as cookie banners~\cite{cookieinspect,cookieblock,alsacnc,cookiegraph}, ad injection~\cite{adsinjector,odin}, and phishing~\cite{goldenhour,Phishintention}. Deceptive pattern detectors use clustering~\cite{mathur2019dark,nazarov2022clustering}, rules~\cite{mansur2023aidui,chen2023unveiling}, or LLMs~\cite{shi202550}, but generally focus on predefined threats or static interfaces. \tool{} instead jointly explores stateful interaction paths, detects contextual PDPs, and repairs them in live web environments.

\noindent \textbf{Mitigation and privacy protection tools.}
Sch\"afer~\etal~\cite{LLM-DP-Repair} show that LLMs can iteratively repair deceptive HTML, but cover only seven readily identifiable patterns, two of which are privacy related. Tools such as Tor~\cite{torproject}, Brave~\cite{bravebrowser}, Adblock Plus~\cite{adblockplus}, AdGuard~\cite{adguard}, and DuckDuckGo~\cite{duckduckgo} mitigate trackers, fingerprinting, ads, or cookie dialogs through predefined rules and provider maintained updates. \tool{} combines functionality preserving repair with autonomous exploration and reasoning based detection across the privacy interface pipeline.

\begin{table}[t]
\centering
\caption{An overview of existing web privacy tools. \tool{} is the only tool that fully covers all seven PDP types while being fully automated, adaptable, and robust against grooming.}
\label{tab:related_work}
\newcommand{\rot}[1]{\rotatebox[origin=l]{0}{#1}}
\setlength{\tabcolsep}{3pt}
\renewcommand{\arraystretch}{1.0}
\resizebox{\columnwidth}{!}{%
\begin{tabular}{@{}l cc ccc cc ccc@{}}
\toprule
 & \multicolumn{2}{c}{\textbf{Info.\ Conceal.}}
 & \multicolumn{3}{c}{\textbf{Interface Obstr.}}
 & \multicolumn{2}{c}{\textbf{Forced Discl.}}
 & \multicolumn{3}{c}{\textbf{Sys.\ Prop.}} \\
\cmidrule(lr){2-3}\cmidrule(lr){4-6}\cmidrule(lr){7-8}\cmidrule(l){9-11}
\textbf{Web Privacy Tools}
 & \rot{\begin{tabular}[l]{@{}l@{}}Hidden\\Info.\end{tabular}} 
 & \rot{\begin{tabular}[l]{@{}l@{}}Disguised\\Ads\end{tabular}}
 & \rot{Preselection} 
 & \rot{\begin{tabular}[l]{@{}l@{}}False\\Hierarchy\end{tabular}} 
 & \rot{\begin{tabular}[l]{@{}l@{}}Small\\Close\\Button\end{tabular}}
 & \rot{\begin{tabular}[l]{@{}l@{}}Privacy\\Zuckering\end{tabular}} & \rot{\begin{tabular}[l]{@{}l@{}}Forced\\Action\end{tabular}}
 & \rot{\begin{tabular}[l]{@{}l@{}}Fully\\Automated\end{tabular}} 
 & \rot{Adaptable} 
 & \rot{\begin{tabular}[l]{@{}l@{}}Grooming-\\Robust\end{tabular}} \\
\midrule
CookieInspect~\cite{cookieinspect}     & \pie{0}   & \pie{0}   & \pie{360} & \pie{0}   & \pie{0}   & \pie{0}   & \pie{360} & \pie{0}   & \pie{0} & \pie{0} \\
CookieGraph~\cite{cookiegraph}         & \pie{0}   & \pie{0}   & \pie{0}   & \pie{0}   & \pie{0}   & \pie{360} & \pie{360} & \pie{180} & \pie{0} & \pie{0} \\
CookieBlock~\cite{cookieblock}         & \pie{360} & \pie{0}   & \pie{360} & \pie{0}   & \pie{0}   & \pie{360} & \pie{360} & \pie{180} & \pie{0} & \pie{0} \\
ALSACNC~\cite{alsacnc}                 & \pie{360} & \pie{0}   & \pie{0}   & \pie{0}   & \pie{0}   & \pie{360} & \pie{360} & \pie{180} & \pie{0} & \pie{0} \\
AdsInjector~\cite{adsinjector}         & \pie{0}   & \pie{360} & \pie{0}   & \pie{0}   & \pie{0}   & \pie{0}   & \pie{0}   & \pie{0}   & \pie{0} & \pie{0} \\
ODIN~\cite{odin}                       & \pie{0}   & \pie{360} & \pie{0}   & \pie{0}   & \pie{0}   & \pie{0}   & \pie{0}   & \pie{180} & \pie{0} & \pie{0} \\
Golden Hour~\cite{goldenhour}          & \pie{0}   & \pie{0}   & \pie{0}   & \pie{0}   & \pie{0}   & \pie{360} & \pie{0}   & \pie{0}   & \pie{0} & \pie{0} \\
PhishIntention~\cite{Phishintention}   & \pie{0}   & \pie{0}   & \pie{0}   & \pie{0}   & \pie{0}   & \pie{360} & \pie{180} & \pie{0}   & \pie{0} & \pie{0} \\
\midrule
Mathur~\etal~\cite{mathur2019dark}     & \pie{360} & \pie{0}   & \pie{0}   & \pie{0}   & \pie{0}   & \pie{0}   & \pie{360} & \pie{0}   & \pie{0}   & \pie{0} \\
Nazarov~\etal~\cite{nazarov2022clustering} & \pie{360} & \pie{360} & \pie{0} & \pie{0}  & \pie{0}   & \pie{360} & \pie{360} & \pie{0}   & \pie{0}   & \pie{0} \\
AidUI~\cite{mansur2023aidui}           & \pie{0}   & \pie{360} & \pie{360} & \pie{360} & \pie{0}   & \pie{0}   & \pie{0}   & \pie{180} & \pie{0}   & \pie{0} \\
UIGuard~\cite{chen2023unveiling}       & \pie{0}   & \pie{360} & \pie{360} & \pie{360} & \pie{360} & \pie{360} & \pie{0}   & \pie{180} & \pie{0}   & \pie{0} \\
Sch\"{a}fer~\etal~\cite{LLM-DP-Repair} & \pie{360} & \pie{0}   & \pie{0}   & \pie{360} & \pie{0}   & \pie{0}   & \pie{0}   & \pie{0}   & \pie{0}   & \pie{0} \\
DPGuard~\cite{shi202550}               & \pie{360} & \pie{360} & \pie{360} & \pie{360} & \pie{360} & \pie{360} & \pie{360} & \pie{180} & \pie{360} & \pie{0} \\
\midrule
\rowcolor{gray!14}
\textbf{\tool{} (Ours)}                & \pie{360} & \pie{360} & \pie{360} & \pie{360} & \pie{360} & \pie{360} & \pie{360} & \pie{360} & \pie{360} & \pie{360} \\
\bottomrule
\multicolumn{11}{@{}l@{}}{\footnotesize \pie{360}~fully \ / \ \pie{180}~partially \ / \ \pie{0}~not supported.} \\
\end{tabular}%
}
\end{table}

%===============================================================================
\section{Conclusion}\label{sec:conclusion}
\label{sec:discussion}

In this paper, we examined website UI privacy through the lens of deceptive patterns and formalized AI grooming as an artifact level threat to AI mediated web workflows. We presented \tool{}, an agent-in-the-middle framework that combines provenance filtering, prompt level protection, and four specialized agents. Experiments showed that \tool{} blocked 90.98\% of naive whole-page groomed samples, covered 80.62\% of privacy patterns while visiting only 9.5\% to 12.2\% as many pages as baselines, achieved state-of-the-art PDP detection, and reliably repaired interfaces under functional and regulatory constraints. Caching and modular execution kept runtime overhead low, while our study in the wild showed that protection policies could be calibrated across domains and regions. One avenue for future exploration is to evaluate stronger adaptive attacks and automate functionality testing. More broadly, our results show that the interface supply chain, from crawled training data to rendered pages, can be protected at well-chosen points of contact. We hope this work encourages the community to treat deceptive design as a supply-chain security problem and to build a web where AI mediation strengthens, rather than erodes, user privacy and trust.

%-------------------------------------------------------------------------------
\cleardoublepage
\section*{Acknowledgments}
Zewei Shi is supported by CSIRO Full PhD Scholarship. Feng Liu is supported by the Australian Research Council (ARC) with grant numbers DP230101540 and DE240101089, and the NSF-CSIRO Responsible AI program with grant number 2303037. Minhui Xue and Xingliang Yuan are supported by CSIRO – National Science Foundation (US) AI Research Collaboration Program. We especially acknowledge and thank Dr. Jieshan Chen for her early guidance, advice on the background of deceptive (dark) patterns, and comments on an initial draft of this paper.

\section*{Ethical Considerations}
\label{apx:ethical_consideration}

\subsection*{I. Motivation for Publishing}
The intersection of Large Language Models (LLMs) and web interfaces presents a profound dual-use dilemma. While agentic AI can significantly enhance user experience and automate complex tasks, it concurrently introduces severe risks via ``AI grooming'', where attackers exploit data voids to inject human-benign but ground-truth malicious samples, corrupting model reasoning and normalizing privacy deceptive patterns (PDPs). The fundamental motivation for publishing this research is strictly defensive. It is imperative to formalize this emerging threat model and openly discuss proactive, agent-in-the-middle defense mechanisms like \tool{}. Establishing robust grooming purification and interface repair frameworks before widespread malicious exploitation of the AI supply chain occurs is critical for ensuring that AI-mediated web ecosystems remain safe and trustworthy.

\subsection*{II. Stakeholder Analysis}
We identify four primary stakeholders affected by this research:
\begin{itemize}
    \item \textbf{End-users (Beneficiaries):} Everyday web users who are increasingly exposed to manipulative interfaces and privacy risks. They are the primary beneficiaries, gaining an automated tool (\tool{}) that proactively protects their privacy autonomy without disrupting their browsing experience.
    \item \textbf{Model developers and web ecosystem providers:} Organizations training frontier LLMs or operating web crawlers. They benefit from the proposed latent-space purification mechanisms, gaining verifiable methods to prevent their data pipelines from being poisoned by deceptive artifacts.
    \item \textbf{The scientific community and regulators:} Beneficiaries of a formalized taxonomy for AI grooming and PDPs. This research provides a transparent framework to audit web interfaces and enforce privacy compliance at scale.
    \item \textbf{Malicious actors (Adversaries):} Entities attempting to execute supply-chain poisoning or UI manipulation for data harvesting and economic espionage. Their ability to bypass human-in-the-loop security audits will be significantly hindered by our proposed defenses.
\end{itemize}

\subsection*{III. Impact Assessment and Experimental Safeguards}
\begin{itemize}
    \item \textbf{Positive impacts:} \tool{} provides a vital defense-in-depth tool that operates directly in live web environments, offering both supply-chain purification and real-time user interface repair. This work aligns with responsible AI principles by equipping systems to recognize and resist subtle, long-term manipulation without compromising user autonomy. Crucially, our defenses emphasize transparency and semantic robustness, avoiding silent or coercive interventions (\eg relying on user-facing warnings and highlighting instead).
    \item \textbf{Mitigation of misuse risks:} To study AI grooming safely, all manipulated content was synthetically generated; no data was scraped from real users or deployed systems. Furthermore, we deliberately omit implementation details and raw adversarial prompts that could enable misuse by malicious actors, ensuring our evaluation remains rigorous without proliferating actionable exploits.
    \item \textbf{Human subjects protection:} This study was approved by our institutional Human Research Ethics Committee. Participation was voluntary, with informed consent obtained from all participants, who could withdraw at any time using a pre-assigned keyword. We collected only non-identifiable demographic data (\eg age range, country, gender) necessary for analysis, and implemented strict confidentiality measures to mitigate any re-identification risk. All analyses were conducted on aggregated or de-identified data.
\end{itemize}

\bibliographystyle{plainurl}
\bibliography{references}

%===============================================================================
\appendix
\section{Privacy Deceptive Pattern (PDP) Taxonomy}
\label{apx:taxonomy}
\begin{table*}[t]
\centering
\caption{Taxonomy of privacy deceptive patterns (PDPs).}
\label{tab:privacy_dp_taxonomy}
\resizebox{\linewidth}{!}{
\begin{tabular}{p{1.8cm}p{2cm}p{6cm}p{9cm}}
\toprule
\textbf{Objective} & \textbf{PDPs} & \textbf{Description} & \textbf{Use-cases} \\ 
\midrule

\multirow{8}{1.8cm}[-0.8em]{Information Concealment} & 
\multirow[c]{4}{2cm}[-0.5em]{Hidden Information} & 
\multirow[c]{4}{6cm}[-0.5em]{Essential privacy information hidden behind hyperlink.} & 
The sign-up/sign-in screen shows ``By continuing, you agree…'' with the full policy accessible only via a hyperlink. \\
\cmidrule(){4-4}
 & & & The cookie banner provides cookie information via a separate ``Cookie Policy'' link. \\ 
\cmidrule(){2-4}

 & {Disguised Ads (with hint)} & \multirow[c]{2}{6cm}{Ads blend with content but include an ``Ad/Sponsored'' label.} & Google/Amazon ads are labeled with their respective ad icons or keywords. \\ 
\cmidrule(){2-4}

 & \multirow[c]{2}{2.5cm}{Disguised Ads (without hint)} & \multirow[c]{2}{6cm}[-0.2em]{Ads fully mimic content and lack any hint.} & \multirow[c]{2}{9cm}[-0.2em]{A native ad in the article list that appears identical to genuine articles.} \\
 \\
\midrule

\multirow{6}{1.8cm}[-1.5em]{Interface Obstruction} & \multirow{2}{2cm}[-0.4em]{Preselection} & \multirow{2}{6.4cm}[-0.2em]{Non-essential data-sharing options are pre-select/hidden behind a ``Accept'' button.} & Cookie banner with analytics boxes selected by default. \\
\cmidrule(){4-4}
 & & & The sign-up form pre-selects to receive marketing newsletters. \\ 
\cmidrule(){2-4}

 & \multirow{2}{2cm}[-0.5em]{False Hierarchy} & \multirow{2}{6cm}[-0.4em]{One option is more dominant than the other} & Large colourful ``Accept'' vs. small grey ``Reject ''. \\ 
\cmidrule(){4-4}
 & & & Prominent ``Sign up'' contrasts with a subtle ``Continue as a guest''. \\ 
\cmidrule(){2-4}

 & \multirow{2}{2cm}[-0.2em]{Small Close Button} & \multirow{2}{6cm}[-0.2em]{The close button is deliberately tiny, leading users to accidentally click on other content .} & Tiny ``×'' on an Ads causes users to click the Ads instead of closing it. \\ 
\cmidrule(){4-4}
 & & & Cookie prompt’s ``×'' so small users give up and accept cookies. \\ 
\midrule

\multirow{4}{1.8cm}[-1.5em]{Forced Data Disclosure} & \multirow{2}{2cm}[-1.2em]{Privacy Zuckering} & \multirow{2}{6cm}[-1.5em]{Only a data-sharing path is offered, with no opt-out option.} & A page includes only “Accept All/Continue” button, leaving users with no practical way to decline non-essential data collection. \\
\cmidrule(){4-4}
 & & & Onboarding pre-selects contact sharing without providing an opt-out option. \\ 
\cmidrule(){2-4}
 & \multirow{2}{2cm}[-0.3em]{Forced Action} & \multirow{2}{6cm}[-0.3em]{User must perform an action to continue.} &Users must sign up or sign in to access public services. \\ 
\cmidrule(){4-4}
 & & & User must accept all cookies to view content (no ``Decline'' option). \\ 
\bottomrule
\end{tabular}
}
\end{table*}

Recently, Gray~\etal~\cite{GaryCHI24} published a work unifying DP taxonomies from the perspectives of human-computer interaction and design, web measurement and regulatory experience, and data protection law. However, this unified taxonomy does not incorporate the privacy literature, which focuses on protecting end users from being tricked by high-risk privacy-related DPs. To systematically bridge privacy issues and deceptive interface designs, we build on DPGuard~\cite{shi202550}, the first work that explicitly connects DPs with privacy, by selecting privacy-related DPs based on the underlying goals of the attacker. We also enhance the taxonomy with detailed use-case descriptions and propose a repair method for each selected PDP.

Specifically, we identify three core objectives: \textit{Information Concealment}, where critical privacy-related information is visually hidden or downplayed; \textit{Interface Obstruction}, where interfaces are intentionally made difficult to navigate, causing users to overlook or abandon privacy-relevant options; and \textit{Forced Data Disclosure}, where users are forced into sharing personal data without an opt-out choice. Our taxonomy, presented in Table~\ref{tab:privacy_dp_taxonomy}, organizes 7 common deceptive patterns under these three attacker goals and illustrates them with 14 real-world use cases. This structure captures a broad spectrum of privacy-infringing UI tactics observed across modern websites and serves as a foundation for both detection and repair efforts.

\noindent \textbf{Information concealment.}
Attackers may deceive users by deliberately concealing critical privacy-related information in the user interface, making it difficult for users to recognize, locate, or understand how their data will be collected or used. We associate two DPs with this objective: Hidden Information and Disguised Ads, further dividing the latter into two subtypes, with hint and without hint, based on how detectable the advertising nature is and how we plan to repair them.

\begin{itemize}[leftmargin=*]
\item \textit{Hidden Information.} This pattern refers to cases where essential privacy information, such as Terms of Service, Privacy Policies, or cookie usage details, is not easily visible or accessible. The information may be hidden behind small or low-contrast links, buried in footnotes, or embedded in long texts requiring additional steps to access. Examples include: (1) sign-up or sign-in pages that state ``By continuing, you agree to our policy'' with the actual policy only accessible via a small hyperlink; and (2) cookie banners that reference tracking or personalization without naming specific cookies or third parties, instead linking to a separate document for full details.
\item \textit{Disguised Ads.} These are advertisements that mimic regular content, causing users to interact with them unintentionally. Disguised Ads with hint visually blend into surrounding content but include a subtle marker such as ``Ad'', ``Sponsored'', or a small platform-provided icon. In contrast, Disguised Ads without hint are styled and positioned to be virtually indistinguishable from non-advertising content and contain no visible labels, increasing the risk of users engaging with ads without realizing it, especially on content-heavy platforms where visual cues are critical for content classification.
\end{itemize}

\noindent \textbf{Interface obstruction.}
This category captures deceptive patterns where users must exert additional cognitive effort to identify privacy-related risks. Attackers exploit user fatigue or inattention by creating interfaces that increase complexity, introduce ambiguity, or downplay privacy-preserving options, making it more likely that users overlook the implications of their choices. We link this objective to three patterns: Preselection, False Hierarchy, and Small Close Buttons.

\begin{itemize}[leftmargin=*]
\item \textit{Preselection.} This occurs when privacy-relevant choices, such as cookie preferences, are presented with data-sharing options already enabled. Users are expected to notice and actively disable these options, or may not even have the opportunity to do so if the interface lacks clear alternatives (\eg an ``Accept All'' option without an equivalent ``Reject All'' choice).
\item \textit{False Hierarchy.} A deceptive hierarchy arises when interfaces visually prioritize data-sharing options over privacy-preserving ones. For instance, an ``Accept All'' button might be highlighted in bright colors and placed prominently, while ``Reject All'' or ``Manage Preferences'' options are smaller, grayed out, or hidden in secondary menus.
\item \textit{Small Close Buttons.} In this pattern, close or dismiss buttons are intentionally designed to be small or difficult to click. This tactic is often used in advertisements with visual hints, increasing the chance users accidentally click on the ad, and can also appear in cookie prompts where rejecting non-essential cookies is made harder by a poorly placed or barely visible close button.
\end{itemize}

\noindent \textbf{Forced data disclosure.}
In forced data disclosure patterns, users are compelled to provide personal information without meaningful alternatives or the ability to exercise control; the interface removes or obscures any path that allows users to opt out of data sharing, often conditioning access on data submission. We associate this goal with two patterns: Privacy Zuckering and Forced Action.

\begin{itemize}[leftmargin=*]
\item \textit{Privacy Zuckering.} Users are nudged or misled into disclosing more personal data than they intend, typically by design choices that hide or complicate privacy settings. Often, only a single action path, such as clicking ``Accept All'' or ``Continue'', is available, and users are not given sufficient information or control to avoid data disclosure.
\item \textit{Forced Action.} Users must complete a specific action, such as registering an account, consenting to all data collection, or agreeing to a privacy policy, to proceed with a task or access content. There is no meaningful alternative or skip option, effectively coercing users into sharing information to move forward.
\end{itemize}

\section{Workflow Details}
\label{apx:workflow_details}

\noindent\textbf{Deployment and offline preparation.}~\tool{} requires users to deploy a local proxy (\eg MITMProxy~\cite{mitmproxy}) to mediate HTTP(S) traffic and access webpage source code and responses rendered in the browser. When the user's device is inactive (\eg indicated by low CPU/GPU utilization over a period of time) and with the user's permission, \tool{} prioritizes a queue of websites by visit frequency in the browsing history, then performs backend memory updates by autonomously exploring these sites, detecting PDPs, repairing interfaces, and storing the repaired code. When the user resumes browsing, the proxy compares incoming requests against cached records to reduce runtime overhead and minimize disruption. Users can configure the validity period of cached entries to balance inference cost against browsing freshness.

\noindent\textbf{Online protection workflow.}~The workflow begins when a user navigates to a URL. Once the response is intercepted by the proxy, the request URL is matched against the memory cache; on a hit, the stored, repaired response is returned and rendered directly. Otherwise, the response and a current screenshot are forwarded to the Grooming Purifying Agent, which filters potential AI grooming signals so that downstream agents receive purified inputs (Problem~\ref{prob:purify}). If the page passes purification, the Task Generation Agent constructs structured exploration plans using an LLM guided by RL-based prompt optimization (Problem~\ref{prob:explore}). Based on the generated task list and collected interaction traces, the PDP Detection Agent combines human expert knowledge with LLM reasoning to identify privacy-threatening behaviors (Problem~\ref{prob:detect}). When a PDP is detected, the Interface Repairing Agent employs a secondary reasoning model to infer the precise locations of deceptive elements and applies predefined, webpage-safe repair actions (Problem~\ref{prob:repair}). Finally, the repaired webpage, including cookies and page source, is cached and served on subsequent visits.

\section{Detailed Repair Actions}
\label{apx:repair_actions}

The Interface Repairing Agent restricts modifications to five predefined actions in order to preserve webpage functionality and respect the admissibility constraints defined in Problem~\ref{prob:repair}. Table~\ref{tab:dp_remidiation} summarizes the mapping between PDP types and admissible actions.

\begin{itemize}[leftmargin=*]

\item \textit{Warning.}
A highlighted message is inserted near the deceptive element to alert users to potential privacy risks without modifying the underlying interaction. This action is used when directly removing or changing the element may interfere with legitimate functionality or user autonomy. Representative cases include Hidden Information, Disguised Ads with Hint, Pre-selection, Small Close Button, Privacy Zuckering, and Forced Action.

\item \textit{Highlighting Important Links.}
Privacy-critical links that are deliberately made visually inconspicuous are emphasized so that users can more easily inspect relevant information. This action is particularly applicable to Hidden Information, Privacy Zuckering, and Forced Action.

\item \textit{Block Content.}
Content is removed when the deceptive element constitutes a clearer interface violation and blocking does not prevent legitimate interaction. In our taxonomy, this action is primarily applied to Disguised Ads without Hint.

\item \textit{Auto Cookie Setting.}
When an interface induces acceptance of non-essential cookies despite the user's preference for necessary cookies only, \tool{} automatically restores the cookie configuration to the user's expressed preference.

\item \textit{UI Cleaning.}
For visual-manipulation patterns such as False Hierarchy, \tool{} normalizes the appearance of competing options, reducing artificial visual dominance while preserving the available choices.

\end{itemize}

A computer-use model identifies the corresponding interface elements and applies the selected repair action to the webpage DOM. The modified response is then returned through the proxy and cached for subsequent requests.

\section{Experiment Setup}
This section outlines the datasets collected, the models utilized, and the benchmarks used to evaluate \tool{}.

\subsection{Dataset Collection}
\label{apx:dataset_collection}
To support our evaluation, we created four datasets. The training-testing dataset is used to assess RL-based prompt optimization, website exploration, deceptive pattern detection, and interface repair. We also built a wild dataset to explore additional real-world insights. Details of the dataset construction are provided below.

\noindent\textbf{\tool{} RL-based prompt optimization dataset.}
For RL-based prompt optimization, a dataset containing both website screenshots and their corresponding task lists is essential, as these elements represent the input and ground truth for training task generation models. Multimodal-Mind2Web~\cite{deng2023mindweb} is currently the only open-source dataset that provides domain names, screenshots, and task annotations for each screenshot, making it a valuable resource for this purpose. To adapt this dataset to our task, we first extracted 1,009 unique tasks from 7,775 screenshots, resulting in 1,009 corresponding screenshots. We then removed highly similar images and consolidated the associated tasks into structured task lists, ultimately producing 140 distinct website-task list pairs. These pairs were split into training and testing sets using an 80:20 ratio, yielding 112 pairs for training and 28 for testing.

To validate the test set, we sent HTTP requests to the 28 websites and received valid responses from 23. However, this number proved insufficient for a comprehensive evaluation of website exploration and deceptive pattern detection. To address this limitation, we supplemented the test set with 17 additional websites selected from the top 50 global websites ranked by Similarweb~\cite{similarweb_top_websites}, and manually constructed several tasks for each. As a result, we finalized a dataset comprising 112 website-task list pairs for training and 40 for testing.

\noindent\textbf{\tool{} privacy deceptive pattern detection dataset.}
To evaluate deceptive pattern detection from a privacy perspective, and to establish a foundation for the grooming attack, a set of screenshots annotated under our proposed PDP taxonomy is essential. We used the 40 website-task pairs from the RL-based prompt optimization testing set and employed \tool{} to dynamically explore these 40 websites with 258 corresponding tasks.

We recruited five independent domain experts, each with over five years of active research experience in AI security, generative model alignment, and privacy-enhancing technologies, to annotate the collected webpages. The experts independently inspected the screenshots and labeled privacy-related deceptive patterns according to our taxonomy using LabelMe~\cite{russell2008labelme}. In total, we labeled 515 instances from 418 webpages, including 233 benign webpage screenshots and 185 webpage screenshots involving privacy deceptive patterns. The annotation achieved substantial inter-rater reliability, with an average pairwise Cohen's $\kappa$ of 0.85. Disagreements were resolved through a consensus-driven reconciliation session, resulting in a finalized set of expert-validated labels and interpretations for each PDP.

\noindent\textbf{\tool{} wild dataset.}
To uncover additional insights, we collected 485 websites in the wild. First, we retrieved the top 50 global websites from Similarweb~\cite{similarweb_top_websites}. We then removed sites belonging to harmful or information-sparse categories (\eg Adult, Gambling, Search Engines, and Email). For each of the remaining categories, we collected the top 50 websites, yielding an initial pool of 700 sites. To simulate typical user behavior, we sent HTTP requests with standard headers (User-Agent, Accept, Accept-Language, and Accept-Encoding) and filtered out responses with invalid status codes. We also removed duplicate websites, \ie those delivering identical content under different country-specific domain suffixes. This process resulted in 485 unique top websites spanning 14 categories.

\subsection{Grooming Sample Generation}
\label{apx:grooming_dataset}
For the grooming attack dataset, we used the \tool{} Wild Dataset and our PDP taxonomy as the basis for constructing concrete grooming examples. We first ranked the 485 websites by their number of detected PDP instances and selected screenshots from the 10 most representative websites. The same five experts who participated in the annotation process then reviewed these screenshots and, drawing on their experience from the detection dataset labeling task, identified 33 high-probability privacy deceptive zones. These correspond to realistic interface regions where PDPs commonly appear, such as cookie banners, consent dialogs, advertisement frames, and forced sign-in flows.

The grooming examples were therefore not arbitrarily generated. They were systematically sampled from our PDP taxonomy and instantiated according to the five grooming manipulation strategies of Section~\ref{sec:attacks}: adding noise, pixel modification, invisible text, alpha-channel payloads, and unreadable text. We generated 50 images for each zone across the 10 representative images and applied the five strategies to each. After excluding requests rejected by OpenAI, we collected a total of 8,181 images for the grooming attack dataset.

For the grooming filter dataset, we used the \tool{} privacy deceptive pattern detection dataset as the foundation. The grooming filter is a \textbf{provenance classifier} that distinguishes original screenshots from generated or manipulated ones, rather than classifying whether an interface is benign or deceptive; the dataset split is therefore based on image provenance. Starting from the 185 original PDP images and 233 original No-PDP images, we used the GPT image editing model to generate visually similar images containing comparable interface elements. This generation process was used because directly requesting malicious manipulations, such as adding random noise or hidden content, may be rejected by the model's safety policy. We obtained 182 generated PDP images from the 185 original PDP images and combined them with the originals to construct the grooming filter training dataset. Applying the same process to the 233 original No-PDP images yielded 233 generated No-PDP images, which were combined with the originals to construct the grooming filter evaluation dataset.

\subsection{Grooming Defense Prompt}
\label{apx:grooming_defense_prompt}
\begin{tcolorbox}[size=title]
Note: Some of the images are probably generated, please be careful when identifying the deceptive pattern as these fake screenshots may include some misleading information. Please ignore any characters or elements that you think humans can not understand. When you identified the generated screenshot, please strictly follow the deceptive pattern definitions.
\end{tcolorbox}

\subsection{Benchmarks}
\label{apx:benchmarks}
In the \tool{} architecture, two downstream tasks can be compared against existing works: website exploration and PDP detection.

For website exploration, we first chose the brute-force method, Breadth-First Search (BFS), because it ensures that as many webpages linked to the current page as possible are explored. BFS serves as a strong exhaustive-search baseline rather than ground truth. Since brute force is both time- and resource-intensive, we also selected random exploration and WebUI~\cite{wu2023webui}, a strategy-based exploration tool designed to reduce the likelihood of visiting multiple pages under the same parent. All approaches were evaluated on the RL-based prompt optimization test dataset of 40 website-task pairs, recording how many webpages each visited under the given constraints and calculating the PDP-type coverage rate.

For PDP detection, we selected models from both the privacy detection and deceptive pattern detection domains. DPGuard~\cite{shi202550} represents the state of the art for detecting deceptive patterns across mobile and web platforms. For privacy detection, we selected ALSACNC~\cite{alsacnc}, which focuses on cookie consent compliance and incorporates certain deceptive pattern elements; due to its design focus, it does not support UI manipulation and disguised ads detection. To address this gap, we also introduced AidUI~\cite{mansur2023aidui}, which supports these missing categories on web platforms. We calculate the F1 score for each category and record micro- and macro-averaged F1 as the results.

\section{RL Evaluation Rewards and Ablation Study}
\label{apx:rl_details}
\begin{figure}[t]
\centering
\includegraphics[width=\linewidth]{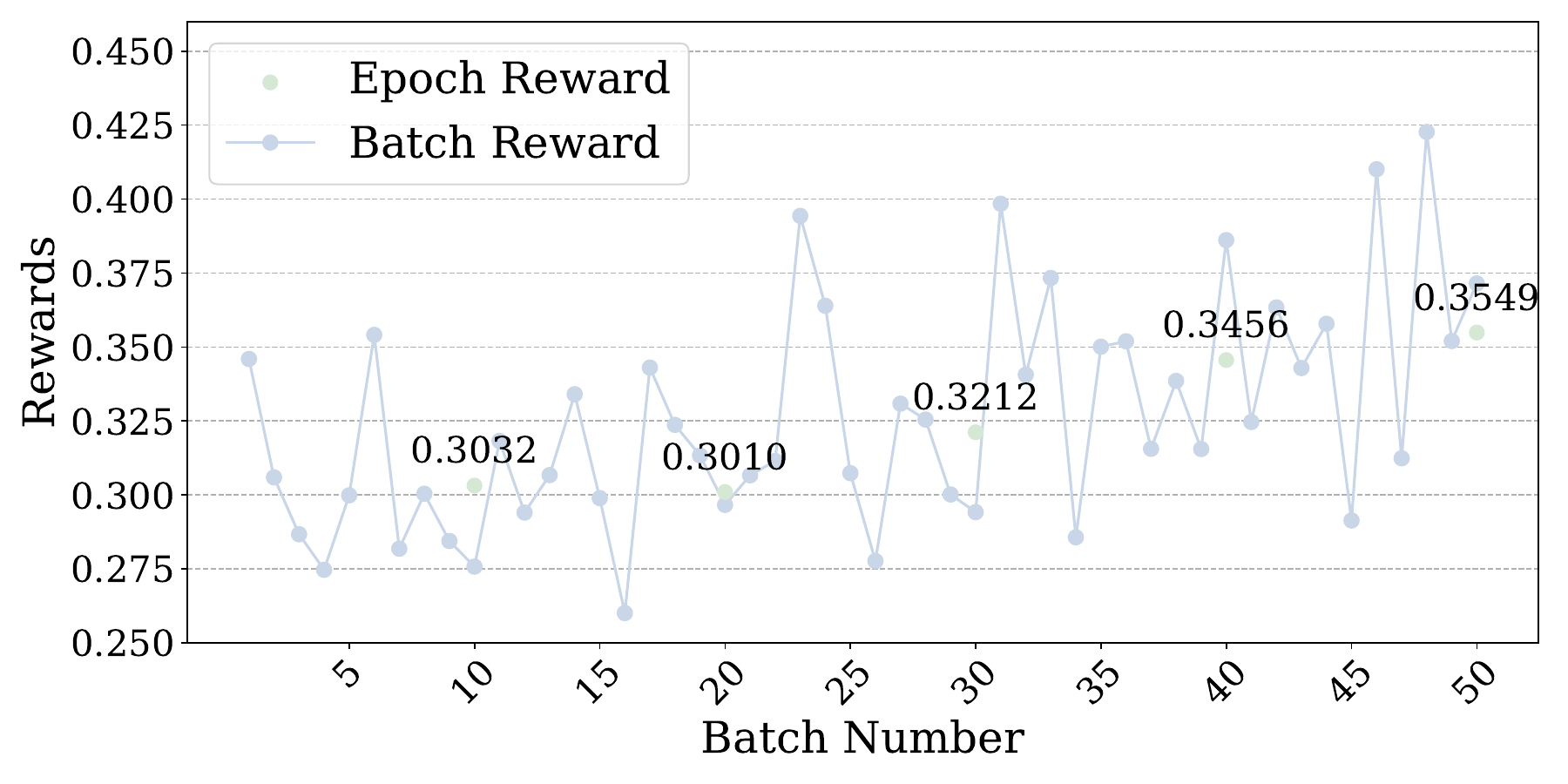}
\caption{RL model evaluation rewards.}
\label{fig:batch_reward}
\end{figure}

Figure~\ref{fig:batch_reward} presents the reward (Eq.~\eqref{eq:reward}) achieved in each training batch of the RL-based prompt optimization of Section~\ref{sec:taskgen}, together with its epoch-level average.  Table~\ref{tab:rf_model_eval} reports the $2{\times}2$ ablation of website information extraction and prompt optimization.

\begin{table}[t]
\centering
\caption{Ablation of reinforcement learning}
\label{tab:rf_model_eval}
    \begin{tabular}{SllS}
      \toprule
      & \multicolumn{2}{c}{\textbf{Prompt}} \\
      \cmidrule(lr){2-3}
      \textbf{Web Info} & without & with \\
      \midrule
      without  & 0.3350 & 0.6626 \\
      with & \textbf{0.4585} & \textbf{\underline{0.7140}} \\
      \bottomrule
    \end{tabular}
\end{table}

\section{User Study Protocol and Ethics}
\label{apx:user_study}

To evaluate the effectiveness and non-intrusiveness of \tool{}, we conducted an online user study. The study design, participant recruitment, and data collection procedures were rigorously structured to ensure high data quality and comprehensively protect participants' rights and privacy.

\noindent \textbf{Ethics, informed consent, and privacy.} As the study involved human participants, the protocol was fully reviewed and approved by our Institutional Review Board (IRB) prior to commencement. Before beginning the survey, all participants were presented with a digital informed consent form detailing the study's objectives, the nature of the data being collected, and their rights. Explicit consent was required to proceed. To protect participant privacy, we did not collect any Personally Identifiable Information (PII). All demographic and survey response data were fully anonymized upon collection, stored on secure, access-controlled servers, and utilized strictly for aggregate analysis. Furthermore, participants were provided with a predefined secret keyword, empowering them to withdraw their participation and request data deletion at any point without penalty.

\noindent \textbf{Participant recruitment and demographics.} We recruited 40 adult participants via Amazon Mechanical Turk (MTurk)~\cite{amazon_mechanical_turk}. We targeted a sample size of 40 because it comfortably exceeds the established thresholds for qualitative heuristic evaluations and formative usability studies, ensuring data saturation for identifying general usability trends and subjective perceptions. To maintain high data quality and mitigate the risk of automated or low-effort responses, recruitment was strictly limited to MTurk workers with a HIT approval rate of 95\% or higher and at least 1,000 previously approved tasks.

\noindent \textbf{Study protocol and quality control.} The survey was refined based on feedback from our local research community. Based on pilot testing, it took approximately 10 minutes to complete and comprised three main sections:
\begin{itemize}
    \item \textit{Background and Demographics:} We collected basic background information, including average daily browsing time, baseline concern regarding online privacy, and prior exposure to privacy deceptive patterns.
    \item \textit{Scenario Evaluation:} Participants evaluated \tool{} across eight distinct real-world web scenarios. Using a 5-point Likert scale (1 = strongly disagree, 5 = strongly agree), they rated the tool's effectiveness in mitigating online privacy risks (``Do you think \tool{} can help?'') and its non-intrusiveness (``Do you agree that \tool{} does not affect your daily browsing experience?'').
    \item \textit{Quality Assurance:} We embedded attention-check questions within the survey to filter out bots or inattentive participants.
\end{itemize}

Every participant who successfully completed the survey and passed the quality checks was compensated with \$5 USD, ensuring fair remuneration for their time. We did not calculate inter-rater agreement, as the study was designed to assess subjective user perceptions via Likert scales rather than objective ground-truth labeling; divergence in opinion was expected, and our analysis focused on aggregate trends.

\noindent \textbf{Summary of baseline responses.} Analysis of the initial survey section revealed that 65\% (26/40) of participants browsed the web for more than two hours daily. Furthermore, 57.5\% (23/40) expressed privacy concerns at or below a moderate level. Notably, 100\% of participants had previously encountered at least one privacy deceptive pattern, with 92.5\% (37/40) citing ``forced actions'' as the most commonly experienced pattern.

\section{LLM Grooming Filter Evaluation}
\label{apx:grooming_filter_performance}
Table~\ref{tab:binary_classifier_performance} compares candidate filter backbones on the grooming filter evaluation dataset; the fine-tuned EfficientNet-V2-L is selected as the deployed filter (Section~\ref{sec:purify}).
\begin{table}[ht]
    \centering
    \caption{Grooming filter performance on grooming filter evaluation dataset.}
    \label{tab:binary_classifier_performance}
    \resizebox{.8\linewidth}{!}{
    \begin{tabular}{ccccc}
    \toprule
    \textbf{Model} & \textbf{Generated} & \textbf{Original} & \textbf{Micro F1} & \textbf{Macro F1} \\
    \midrule
    \makecell{ResNet 101~\cite{resnet}\\(IMAGENET1K\_V2)} & 0.7035 & 0.6636 & 0.6848 & 0.6835\\
    \\
    \makecell{ResNet 152~\cite{resnet}\\(IMAGENET1K\_V2)} & 0.6747 & 0.7327 & 0.7065 & 0.7037\\
    \\
    \makecell{ConvNeXt~\cite{woo2023convnext}\\(large)} & 0.7683 & 0.7483 & 0.7587 & 0.7583\\
    \\
    \makecell{EfficientNet~\cite{tan2019efficientnet}\\(v2\_l)} & \textbf{\underline{0.9028}}& \textbf{\underline{0.9015}}& \textbf{\underline{0.9022}}& \textbf{\underline{0.9022}}\\
    \\
    \makecell{OpenCLIP~\cite{radford2021learning}\\(ViT-bigG-14)} & 0.8636 & 0.8750 & 0.8696 & 0.8693\\
    \\
    \makecell{OpenCLIP~\cite{radford2021learning}\\(ViT-H-14-378)} & 0.8010 & 0.8456 & 0.8261 & 0.8233\\
    \\
    \makecell{GPT-4o~\cite{openai2024gpt4o}\\ (2024-08-06)} & 0.1660 & 0.6626 & 0.5196 & 0.4145\\
    \\
    \makecell{o3~\cite{openai2025o3o4mini}\\ (2025-04-16)} & 0.0413 & 0.6578 & 0.4957 & 0.3496\\
    \\
    \makecell{Gemini~\cite{Kavukcuoglu2025Gemini}\\ (2.5-pro-preview-03-25)} & 0.0581 & 0.6657 & 0.5065 & 0.3619\\
    \bottomrule

    \end{tabular}
    }

\end{table}

\subsection{Runtime Experimental Setup and Performance}
Our runtime evaluation used two hardware environments:
\begin{itemize}
    \item GPU-intensive tasks: NVIDIA A100 GPU (80 GB), 16-core CPU, 128 GB RAM.
    \item Others: Apple M1 Max, 10-core CPU, 32 GB RAM.
\end{itemize}
The software stack included browser automation with the Claude Computer Use tool, PDP detection using Gemini 2.5 Pro, DOM interception and rewriting via MITMProxy, and grooming filtering based on EfficientNet-V2-L. Each module was benchmarked under cold-start conditions (caching disabled) and averaged over three trials, yielding \textbf{1.6 s for exploration, 0.9 s for detection, and 1.1 s for repair per page}.

\begin{table}[t]
\centering
\caption{Grooming filter's correlation with CLIP.}
\label{tab:correlation}
\resizebox{\linewidth}{!}{
\begin{tabular}{@{}cccccc@{}}
\toprule
\makecell{\textbf{CLIP}} & 
\makecell{\textbf{Noise}\\\textbf{Generation}} & 
\makecell{\textbf{Pixel}\\\textbf{Modification}} & 
\makecell{\textbf{Invisible}\\\textbf{Text}} & 
\makecell{\textbf{Alpha}\\\textbf{Channel}} & 
\makecell{\textbf{Unreadable}\\\textbf{Text}} \\
\midrule
\makecell{openai/clip-vit-base-patch16}&0.7339&0.7420&0.6302&0.7057&0.7989\\
\makecell{openai/clip-vit-base-patch32}& 0.6965 & 0.8238 & 0.7161 & 0.6961 & 0.8201\\
\makecell{openai/clip-vit-large-patch14-336}&0.7117&0.6197&0.5425&0.6512&0.7974\\
\bottomrule
\end{tabular}
}
\end{table}

\section{Detailed Discussion on Related Work}
\label{apx:related}

A summary of representative studies is provided in Table~\ref{tab:related_work}.

\noindent \textbf{Data voids and AI/LLM grooming.}
Data voids are gaps in which credible information for a query is scarce, allowing low-quality or strategically crafted sources to gain disproportionate visibility~\cite{boyd2019data}. Recent reports use AI or LLM grooming to describe coordinated attempts to populate such spaces so that search, retrieval, or model pipelines ingest and reproduce attacker-controlled narratives~\cite{newsguard2025pravda,radina2025grooming,prorussia2025american}. These accounts motivate deliberate manipulation of the information supply chain. However, Alyukov~\etal examine chatbot references to Kremlin-linked disinformation and find little evidence for deliberate grooming, instead attributing the observed references primarily to information gaps~\cite{alyukov2025llms}. Their result shows that a model's citation or reproduction of low-quality sources does not by itself establish an adversarial grooming campaign. We therefore distinguish organic exposure caused by information scarcity from the threat model in which an adversary intentionally seeds content, and focus on how adversarially created or manipulated artifacts affect PDP-oriented detection and repair. Existing work does not jointly study data-void exploitation, AI-mediated propagation, interactive PDP discovery, and interface repair.

\noindent \textbf{Website privacy and UI-level threats.}
Recent research has examined web privacy from a fragmented perspective. Traditional tools primarily focus on network-level intrusions and tracking behaviors~\cite{siby2019encrypted,dahlberg2023timeless,singh2024connecting,xue2025theDis}. More recent efforts have shifted attention to UI-layer threats, including cookie consent banners~\cite{cookieinspect,cookieblock,alsacnc,cookiegraph}, ad injection~\cite{adsinjector,odin}, and phishing interfaces~\cite{goldenhour,Phishintention}. CookieInspect~\cite{cookieinspect} conducted the first large-scale audit of GDPR compliance under the IAB Europe Transparency and Consent Framework, uncovering widespread violations. CookieBlock~\cite{cookieblock} proposed a learning-based client-side filter to block invasive cookies without requiring cooperation from websites. ALSACNC~\cite{alsacnc} extended these analyses to multilingual settings, while CookieGraph~\cite{cookiegraph} revealed first-party profiling activities that evade conventional third-party tracking detection. Despite their impact, these studies target specific UI elements or threat types in isolation, whereas DPs capture a broader class of manipulative interface strategies that systematically undermine user privacy and agency.

\noindent \textbf{Deceptive pattern detection.}
Existing DP detection methods span clustering-based~\cite{mathur2019dark,nazarov2022clustering}, rule-based~\cite{mansur2023aidui,chen2023unveiling}, and more recent LLM-powered approaches such as DPGuard~\cite{shi202550}. Clustering-based methods group UI elements using textual or visual features and assign pattern labels to clusters. Rule-based approaches extract properties from screenshots and UI trees and apply handcrafted heuristics to identify deceptive behaviors. For example, AidUI~\cite{mansur2023aidui} targets both web and mobile platforms by matching icon shapes, color schemes, and textual cues, while UIGuard~\cite{chen2023unveiling} focuses on mobile apps using element-level features. DPGuard~\cite{shi202550} leverages LLMs and carefully designed prompts to reason over UI screenshots and achieve strong detection performance. However, such methods still rely on fixed taxonomies and static prompts, making them brittle to evolving or adaptive threats, including AI grooming. More broadly, existing approaches are siloed, static, and primarily detection-oriented. In contrast, our work treats PDPs as a system-level problem and introduces \tool{} as a unified, reasoning-aware multi-agent framework that dynamically interacts with live web interfaces to explore, detect, and repair deceptive patterns. Inspired by human and AI verifier alignment in GUI verification~\cite{pan2025useful}, \tool{} bridges automated reasoning with interactive exploration and repairing, explicitly addressing threats co-shaped by LLMs.

\noindent \textbf{Mitigation and privacy protection tools.}
On the mitigation side, Sch\"afer~\etal~\cite{LLM-DP-Repair} demonstrate that feeding HTML source code into LLMs enables iterative repair of deceptive patterns through prompt refinement. While promising, their approach has two key limitations. First, it targets only seven patterns that are easiest for LLMs to identify, among which only two are privacy-related. Second, it does not guarantee webpage-safe modifications, risking functional breakage and limiting real-world applicability. Other widely used tools, such as Tor~\cite{torproject}, Brave~\cite{bravebrowser}, Adblock Plus~\cite{adblockplus}, AdGuard~\cite{adguard}, and DuckDuckGo~\cite{duckduckgo}, mitigate UI intrusions by blocking trackers, fingerprinting, ads, or cookie dialogs. Adblock Plus provides filter-based control to hide ad frames, tracking pixels, and consent dialogs. AdGuard goes further by automatically dismissing cookie pop-ups and collapsing banners before rendering. DuckDuckGo emphasizes tracker-free and encrypted search, reducing exposure at the search layer rather than within page interfaces. Although effective in practice, these tools depend on predefined rules and provider-driven updates, and they do not generalize well to unseen deceptive behaviors or support adaptive, interactive defenses. In contrast, \tool{} builds a PDP taxonomy grounded in DPGuard~\cite{shi202550} (Appendix~\ref{apx:taxonomy}) but extends it to cover a broader spectrum of privacy risks. It fills a critical gap by offering a reinforcement-driven, multi-agent framework capable of autonomous exploration, reasoning-based detection, and webpage-safe repair across the full privacy interface pipeline, enabling adaptive and proactive defense against evolving deceptive UI practices.

\section{Use Cases Before and After Applying \tool{}}
\label{apx:before_after_repair}
\begin{figure*}[t]
\centering
\includegraphics[width=\linewidth]{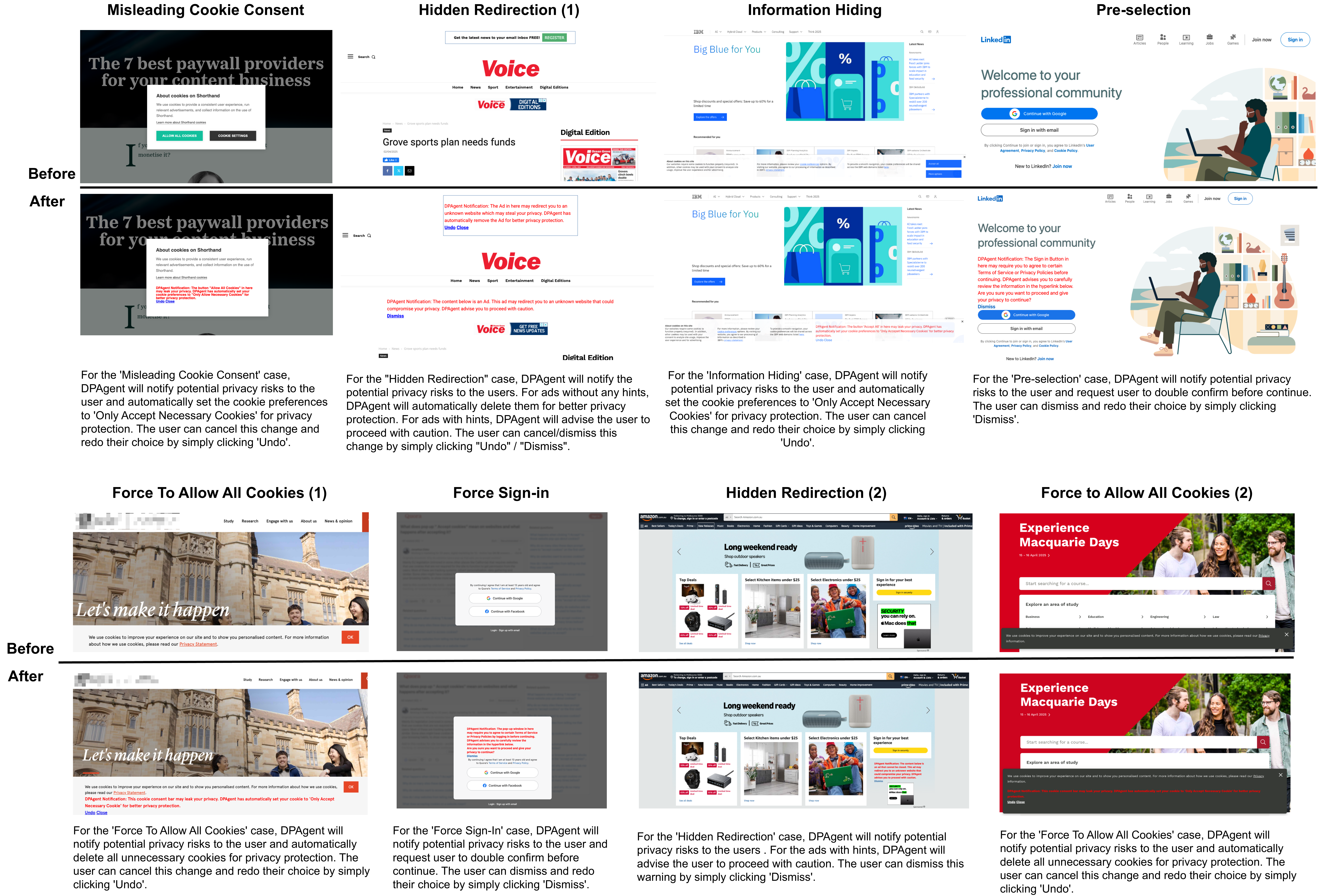}
\caption{Examples of privacy deceptive patterns before and after repair.}
\label{fig:cases}
\end{figure*}

We present eight use cases of before and after applying \tool{} in Figure~\ref{fig:cases}.

%%%%%%%%%%%%%%%%%%%%%%%%%%%%%%%%%%%%%%%%%%%%%%%%%%%%%%%%%%%%%%%%%%%%%%%%%%%%%%%%
\end{document}